\documentclass[iicol]{sn-jnl}

\usepackage{latexsym}
\usepackage[english]{babel}
\usepackage{graphicx}
\usepackage[utf8]{inputenc} 
\usepackage{xcolor}
\usepackage{soul}
\usepackage{amsmath}
\usepackage{amssymb}
\usepackage{amsfonts}
\usepackage{textcomp}
\usepackage{txfonts}
\usepackage{lscape} 
\usepackage{tabularx} 
\usepackage{here}
\usepackage{siunitx}
\DeclareSIUnit\bar{bar}
\usepackage{nicefrac}
\usepackage{enumerate}
\usepackage[numbers,sort&compress]{natbib}
\usepackage{hyperref}
\usepackage{blindtext}
\usepackage{textcomp}
\usepackage{subcaption}

\usepackage{url}

\begin{document}

\title{A cryogenic Paul trap for probing the nuclear isomeric excited state $^{229\text{m}}$Th$^{3+}$}

\author[1]{\fnm{Daniel} \sur{Moritz}}
\equalcont{These authors contributed equally.}

\author[1]{\fnm{Kevin} \sur{Scharl}}
\equalcont{These authors contributed equally.}

\author*[1]{\fnm{Markus} \sur{Wiesinger}}\email{markus.wiesinger@lmu.de}
\equalcont{These authors contributed equally.}

\author[1]{\fnm{Georg} \sur{Holthoff}}

\author[1]{\fnm{Tamila}\sur{Teschler}}

\author[1,a]{\fnm{Mahmood I.} \sur{Hussain}}

\author[2]{\fnm{Jos\'e R.} \sur{Crespo L\'opez-Urrutia}}

\author[3,4]{\fnm{Timo} \sur{Dickel}}

\author[1,b]{\fnm{Shiqian} \sur{Ding}}

\author[4,5,6]{\fnm{Christoph E.} \sur{Düllmann}}

\author[7]{\fnm{Eric R.} \sur{Hudson}}

\author[1,c]{\fnm{Sandro} \sur{Kraemer}}

\author[1,d]{\fnm{Lilli} \sur{Löbell}}

\author[5,6]{\fnm{Christoph} \sur{Mokry}}

\author[4,5]{\fnm{Jörg} \sur{Runke}}

\author[1,e]{\fnm{Benedict} \sur{Seiferle}}

\author[1,f]{\fnm{Lars} \sur{von der Wense}}

\author[1,f]{\fnm{Florian} \sur{Zacherl}}

\author[1]{\fnm{Peter G.} \sur{Thirolf}}

\affil[1]{\orgdiv{LMU München}, \orgname{Fakultät für Physik}, \orgaddress{\city{Garching bei München}, \postcode{85748}, \country{Germany}}}

\affil[2]{\orgdiv{MPIK Heidelberg}, \orgaddress{\city{Heidelberg}, \postcode{69117}, \country{Germany}}}

\affil[3]{\orgdiv{Justus-Liebig-Universität Gießen}, \orgname{II. Physikalisches Institut}, \orgaddress{\city{Gießen}, \postcode{35392}, \country{Germany}}}

\affil[4]{\orgdiv{GSI Helmholtzzentrum für Schwerionenforschung GmbH}, \orgaddress{\city{Darmstadt}, \postcode{64291}, \country{Germany}}}

\affil[5]{\orgdiv{Johannes Gutenberg-Universität}, \orgname{Department of Chemistry - TRIGA site}, \orgaddress{\city{Mainz}, \postcode{55099}, \country{Germany}}}

\affil[6]{\orgdiv{Helmholtz Institute Mainz}, \orgaddress{\city{Mainz}, \postcode{55099}, \country{Germany}}}

\affil[7]{\orgdiv{UCLA}, \orgname{Department of Physics and Astronomy}, \orgaddress{\city{Los Angeles}, \postcode{90095}, \state{California}, \country{USA}}}

\affil[a]{\textit{Present Address:} \orgdiv{Qatar Center for Quantum Computing, College of Science and Engineering, Hamad Bin Khalifa University}, \orgaddress{\city{Doha}, \country{Qatar}}}

\affil[b]{\textit{Present Address:} \orgdiv{Tsinghua University}, \orgname{Department of Physics}, \orgaddress{\city{Beijing}, \postcode{100084}, \country{China}}}

\affil[c]{\textit{Present Address:} \orgdiv{KU Leuven}, \orgname{Institute for Nuclear and Rad. Physics}, \orgaddress{\city{Leuven}, \postcode{3001}, \country{Belgium}}}

\affil[d]{\textit{Present Address:} \orgdiv{Technische Universität München}, \orgname{TUM School of Natural Sciences, Department of Physics}, \orgaddress{\city{Garching bei München}, \postcode{85748}, \country{Germany}}}

\affil[e]{\textit{Present Address:} \orgdiv{Europäisches Patentamt}, \orgaddress{\city{München}, \postcode{80469}, \country{Germany}}}

\affil[f]{\textit{Present Address:} \orgdiv{Johannes Gutenberg-Universität}, \orgname{Institut für Physik}, \orgaddress{\city{Mainz}, \postcode{55128}, \country{Germany}}}

\date{Received: date / Revised version: date}

\abstract{
While laser excitation of the nuclear isomeric transition in $^{229}$Th has been recently achieved for thorium atoms embedded in large-bandgap crystals, laser excitation and characterization of the nuclear transition in trapped $^{229}$Th$^{3+}$ ions has not yet been accomplished. To address these experiments, a cryogenic Paul trap setup has been designed, built, and commissioned at LMU Munich. Here, we present the specifications of the new experimental platform and demonstrate its successful operation, showing the extraction, subsequent ion-guiding, mass-purification, and trapping of $^{229}$Th$^{3+}$ and $^{229\text{m}}$Th$^{3+}$ ions from a newly designed buffer-gas stopping cell as well as of $^{88}$Sr$^{+}$ ions from laser ablation of a solid target. Further, we show sympathetic laser cooling of $^{229\text{(m)}}$Th$^{3+}$ by Doppler-cooled $^{88}$Sr$^{+}$ ions and the formation of mixed-species Coulomb crystals.
}

\maketitle

\section{Introduction}
Among the presently known about 3400 nuclides with a total of more than 186000 nuclear levels \cite{NNDCnudat}, the first isomeric excited state of \textsuperscript{229}Th, denoted as \textsuperscript{229m}Th, has the lowest known excitation energy. At about \SI{8.4}{\electronvolt}, it is in the range of outer-shell electronic transitions, providing accessibility to the isomeric state with current laser technology, thus resulting in a unique standing of \textsuperscript{229m}Th among all presently known nuclei. The existence of the isomeric first excited state \textsuperscript{229m}Th as an almost degenerate ground-state doublet has already been conjectured from $\gamma$ spectroscopic data by Kroger and Reich in 1976 \cite{Kroger76}. However, it took until 2016 for the first direct detection of the ground-state decay of \textsuperscript{229m}Th via the internal conversion decay channel \cite{Nature16Wense} and until 2023 for the first observation of the isomer's radiative decay via vacuum-ultraviolet (VUV) spectroscopy of \textsuperscript{229m}Th embedded in large-bandgap crystals (CaF\textsubscript{2} and MgF\textsubscript{2}) at the ISOLDE facility \cite{Nature23Kraemer}. 

Recently, the first successful laser excitation of the isomer embedded in CaF\textsubscript{2} has been reported by PTB in cooperation with TU Vienna \cite{PhysRevTiedauSchaden24} followed by the laser excitation of \textsuperscript{229m}Th embedded in LiSrAlF\textsubscript{6} reported from UCLA \cite{elwellPhysRevLett24}. Shortly thereafter, spectroscopy with a VUV frequency comb based on high harmonic generation (HHG) allowed the determination of the \textsuperscript{229m}Th transition with so far unprecedented precision, resolving the electric quadrupole splitting of the transition and resulting in a value of $\nu_{\text{Th}}=\SI{2020407384335(2)}{\kilo\hertz}$ for the unsplit transition \cite{zhang2024_nature}. This was achieved by establishing a direct frequency link between the JILA \textsuperscript{87}Sr lattice clock and the VUV frequency comb brought into resonance with the isomeric transition in \textsuperscript{229}Th embedded in CaF\textsubscript{2}.
Very recently, also laser excitation in ThF\textsubscript{4} was observed \cite{Zhang2024ThF}.

With the radiative half-life of \textsuperscript{229m}Th being on the order of $\SI{e3}{\second}$, as theoretically predicted \cite{PhyRevC15Tkalya} and recently confirmed by measurements \cite{PhysRevTiedauSchaden24,NatureYamaguchi24,elwellPhysRevLett24,zhang2024_nature,Zhang2024ThF}, the resulting natural linewidth of the transition is on the order of $\Delta E/E \approx 10^{-20}$. In addition, the nuclear transition provides low susceptibility to external electromagnetic perturbations because of the small nuclear moments. Due to these properties, \textsuperscript{229m}Th provides the unique opportunity to be used as a nuclear frequency standard, a 'nuclear clock'. While first hints at a possible application of \textsuperscript{229m}Th as frequency standard were already mentioned by Tkalya et al. \cite{PhysScr96Tkalya}, the first viable concept for the implementation of a nuclear frequency standard based on \textsuperscript{229m}Th was proposed in 2003 by Peik and Tamm \cite{PhysLett03Peik}. Following the approach of an optical clock based on \textsuperscript{229}Th\textsuperscript{3+} ions stored and laser cooled in a linear Paul trap, a total systematic uncertainty of 1.5$\times10^{-19}$ might be achieved \cite{PhysRev12Campbell}, thus competing with and even surpassing the currently best existing optical atomic clocks \cite{Brewer19,Aeppli2024}. Such a thorium-based nuclear clock would offer a broad range of applications with perspectives in applied as well as fundamental physics, as outlined in more detail in \cite{AnnPhys19Thirolf,peik2021nuclear}. These applications range from improved precision of satellite-based navigation systems to relativistic geodesy to the use of a nuclear clock as a unique type of quantum sensor in fundamental physics such as the search for ultra-light dark matter candidates and the investigation of theoretically predicted variations of fundamental constants, particularly the fine structure constant $\alpha$ \cite{Flambaum06,Berengut09,Beeks2024}.

Complementing the considerable progress achieved in recent years on the identification and characterization of the thorium isomer \cite{ThirolfEPJ24,PhysRevTiedauSchaden24,NatureYamaguchi24,Zhang2024ThF,elwellPhysRevLett24,zhang2024_nature}, a consolidation of our knowledge on fundamental properties of \textsuperscript{229m}Th now has to focus on the spectroscopy of the nuclear transition in a vacuum environment as spectroscopy has so far only been achieved for \textsuperscript{229}Th embedded in large-bandgap crystals \cite{PhysRevTiedauSchaden24,elwellPhysRevLett24,zhang2024_nature,Zhang2024ThF}. In particular, the determination of the radiative lifetime $\tau$ of \textsuperscript{229m}Th under optimized vacuum conditions has not yet been achieved, as the best currently available result for the half-life of trapped \textsuperscript{229m}Th\textsuperscript{3+} ions of $1400^{+600}_{-300}\,\si{\second}$ (corresponding to a lifetime of $\tau=2020^{+866}_{-433}\,\si{\second}$) \cite{NatureYamaguchi24} was measured at a helium buffer-gas pressure of \SI{0.02}{\milli\bar}. Other recently reported results for the radiative lifetime were measured in large-bandgap crystals and accordingly scaled with $n^3$, where $n$ denotes the refractive index of the medium. For CaF\textsubscript{2} as host material, this resulted in $\tau=\SI{2513(60)}{\second}$ \cite{PhysRevTiedauSchaden24}, $\tau=\SI{2557(16)}{\second}$ \cite{zhang2024_nature}, or $\tau=2589(92)_{stat}(115)_{sys}\,\si{\second}$ \cite{hiraki2024}. In MgF\textsubscript{2} the scaled vacuum lifetime is $\tau=\SI{3189(491)}{\second}$ \cite{Nature23Kraemer}, while for LiSrAlF\textsubscript{6} as host material $\tau=1860(43)_{stat}(66)_{sys}\,\si{\second}$ \cite{elwellPhysRevLett24} has been reported. In ThF\textsubscript{4} thin films, a significantly reduced lifetime is observed which cannot fully be explained by the refractive index of ThF\textsubscript{4} bulk material \cite{Zhang2024ThF}.

In this paper, we describe a cryogenic linear Paul-trap setup which has recently been designed, built, and commissioned at LMU Munich.
This provides the experimental prerequisites for the determination of the radiative lifetime under ultra-high vacuum conditions, as well as to serve as a platform for the spectroscopy of the nuclear transition for trapped ions and therefore as backbone for the nuclear clock prototype to be set up at LMU in the framework of the 'ThoriumNuclearClock' project \cite{ThoriumNuclearClock}. In order to achieve the required long storage times, the linear Paul trap is operated at cryogenic temperatures and the \textsuperscript{229(m)}Th\textsuperscript{3+} ions will be sympathetically cooled by laser-cooled, co-trapped \textsuperscript{88}Sr\textsuperscript{+} ions. 
Therefore, the experimental platform has to enable optical access to the central region of the cryogenic linear Paul trap for the cooling and spectroscopy lasers, for fluorescence detection, as well as for ion-optical beam line components to provide \textsuperscript{229(m)}Th\textsuperscript{3+} and \textsuperscript{88}Sr\textsuperscript{+} ions from their respective sources to the trap. 

In section 2 we provide a detailed description of the experimental apparatus and its components including the ion sources, the ion injection line on one end of the central cryogenic Paul trap, the ion diagnostics line on the other end of the trap, as well as the vacuum system. In addition, we provide a detailed characterization of all components. In section 3 we describe experiments on thorium extraction from the buffer-gas cell. In section 4 we show thorium transport to and trapping in the cryogenic Paul trap, as well as sympathetic cooling of \textsuperscript{229(m)}Th\textsuperscript{3+} by laser-cooled \textsuperscript{88}Sr\textsuperscript{+} ions and the formation of mixed-species Coulomb crystals, and in section 5 we give an estimate of the pressure in the cryogenic ion trap based on the lifetime of laser-cooled \textsuperscript{88}Sr\textsuperscript{+} ions. Finally, we present our conclusions in section 6.

\section{Experimental apparatus}
A cryogenic trap setup has been chosen due to the requirement of long storage times. The design and dimensions of the 4-rod quadrupole ion trap are based on the CryPTEx trap operated in Heidelberg \cite{schwarz2012cryogenic, leopold2019cryogenic}. The required cooling of the cryogenic ion trap is provided by a pulse-tube cryocooler (\textit{RP-082B2-F70H}, Sumitomo Heavy Industries). Crucially, vibration isolation is achieved by a commercial vibration-isolation system \cite{ColdEdgeVibrationInterface} isolating the cryocooler from the trap chamber by a helium atmosphere with high thermal conductivity. Previously, this system was used, e.g., in quantum computing experiments \cite{dubielzig2021ultra,Dubielzig_Thesis} where suppression of vibrations to below \SI{10}{\nano\meter} was achieved. This allows maintenance-free operation of the experiment for several months at a time.

\begin{figure*}[t]
	\centering
	\includegraphics[width = \textwidth]{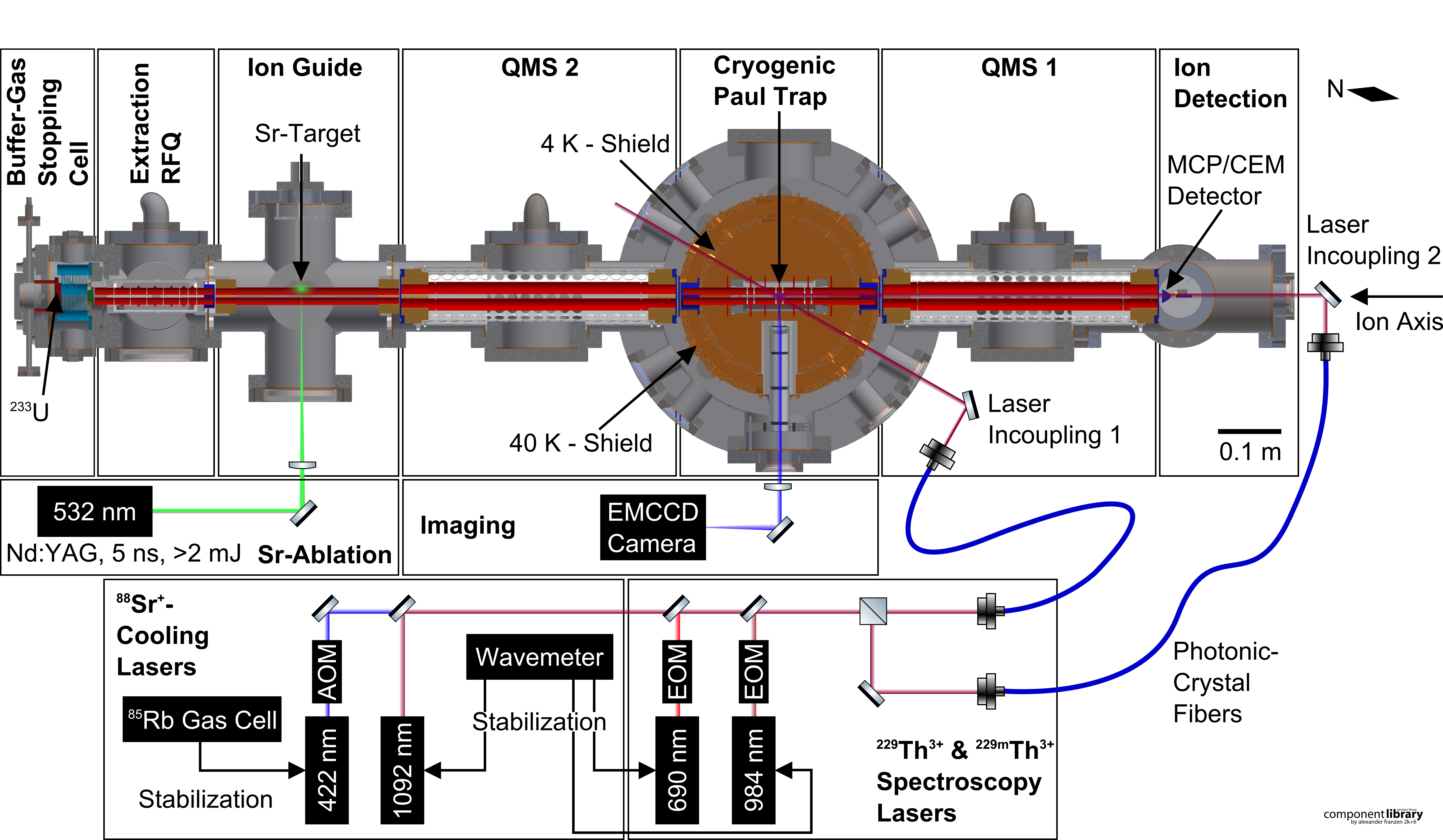}
	\caption{Overview of the cryogenic linear Paul trap setup at LMU: Shown is a horizontal cross-section of the experimental apparatus along the ion axis together with a schematic of the laser setup. The needle, located at the top right, indicates the orientation of Earth's magnetic field. RFQ: radio-frequency quadrupole, QMS: quadrupole mass separator, MCP: micro-channel plate, CEM: channel electron multiplier, EMCCD: electron-multiplying charge-coupled device, AOM: acousto-optic modulator, EOM: electro-optic modulator.
    \label{fig:setupoverview}}
\end{figure*}

The experimental setup features a horizontal ion beam line which is depicted in Fig.~\ref{fig:setupoverview}. Its components are described in detail in the following subsections, and consist of:
\begin{enumerate}
    \item A helium buffer-gas stopping cell housing a thin \textsuperscript{233}U source emitting \textsuperscript{229(m)}Th ions, a radio-frequency (RF) funnel for collimation, and a supersonic de Laval nozzle for extraction;
    \item A segmented RF quadrupole for phase-space cooling and bunching of the ions extracted from the buffer-gas stopping cell (Extraction RFQ);
    \item An Ion Guide chamber containing an RF quadrupole for guiding the \textsuperscript{229(m)}Th ions extracted from the buffer-gas cell as well as the \textsuperscript{88}Sr ions laser ablated from a strontium titanate (SrTiO\textsubscript{3}) target mounted in this chamber;
    \item A quadrupole mass separator (QMS) used to select the ion species and charge state to be loaded into the trap;
    \item A cryogenic linear Paul trap surrounded by oxygen-free-copper heat shields and cooled down to about \SI{8}{\kelvin} by a vibration-isolated cold head forming the centerpiece of the setup; 
    \item A multi-channel plate (MCP) detector, or optionally a channel electron multiplier (CEM) detector, mounted after a second QMS on the other end of the trap which allows for mass-selective ion detection for diagnostic purposes;
    \item A vacuum system consisting of several independently pumped vacuum chambers. In combination with small apertures, this allows for differential pumping of the setup, thus enabling buffer-gas pressures up to \SI{40}{\milli\bar} in the buffer-gas cell while maintaining ultra-high vacuum (UHV) in the trap chamber.
\end{enumerate}
Furthermore, our experimental setup features an imaging system based on aspheric lenses with a collection efficiency of $\approx\SI{1.5}{\percent}$ and an electron-multiplying charge-coupled-device (EM-CCD) camera for detection of the fluorescence photons from laser-cooled \textsuperscript{88}Sr\textsuperscript{+} ions near \SI{422}{\nano\meter} and the fluorescence photons from \textsuperscript{229(m)}Th\textsuperscript{3+} ions near \SI{690}{\nano\meter}, see Fig.~\ref{fig:setupoverview}. Additionally, there is a laser system based on commercial external cavity diode lasers (ECDL) (\textit{DL pro}, Toptica) for laser cooling of \textsuperscript{88}Sr\textsuperscript{+} and laser spectroscopy of electronic transitions in \textsuperscript{229(m)}Th\textsuperscript{3+}. The imaging system and laser setup were already subject of a previous publication \cite{ScharlSetup2023}, to which the reader is referred for a detailed description of those components.

\subsection{The buffer-gas stopping cell}
\label{section:Buffergasstopp}

\begin{figure} [htb]
	\centering
	\includegraphics[width=\linewidth]{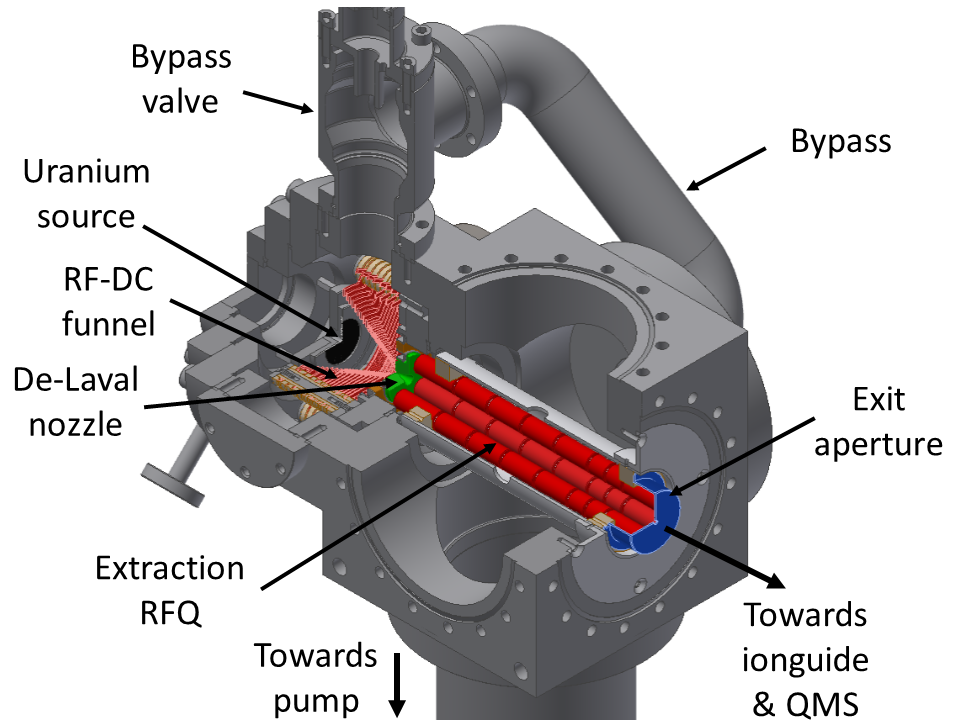}
	\caption{Section view of the buffer-gas stopping cell and the subsequent Extraction RFQ. The vacuum chamber of the RFQ is a modified CF100 cube (side length \SI{152}{\milli\meter}). Feedthroughs as well as pressure gauges attached to the chamber are omitted.}
	\label{fig:stoppcelloverview}
\end{figure}

The design and characterization of the pre-existing buffer-gas stopping cells operated at LMU is comprehensively treated in the pioneering works of Jürgen Neumayr \textit{et al.} \cite{NeumayrThesis,Neumayr06}. For the purpose of loading our ion trap with \textsuperscript{229(m)}Th\textsuperscript{3+} ions, a considerably more compact buffer-gas stopping cell has been set up, see Fig.~\ref{fig:stoppcelloverview}.

\begin{figure}[htb]
		\centering
		\includegraphics[width=\linewidth]{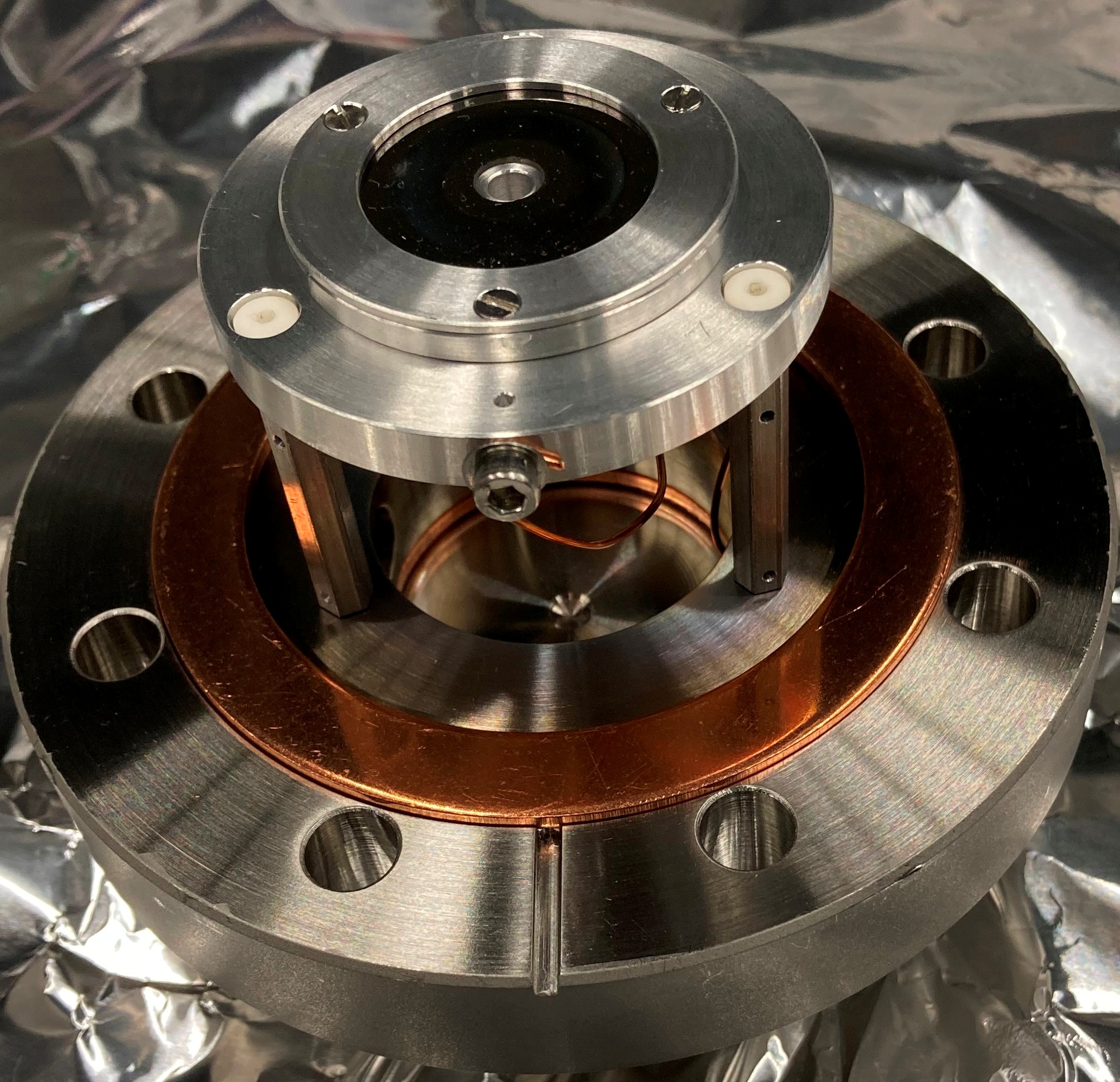}
		\caption{Photograph of the \textsuperscript{233}U source mounted on its CF63 carrier flange.}
		\label{fig:uraniumsource}
\end{figure}

The \textsuperscript{229}Th ions are generated via the $\alpha$ decay of \textsuperscript{233}U with a branching ratio of $\approx \SI{2}{\percent}$ for the isomeric state \cite{Nature16Wense, Thielking18,Wense18} using a \SI{10}{\kilo\becquerel} \textsuperscript{233}U source. This source was fabricated at the TRIGA site of the department of chemistry at the University of Mainz \cite{Eberhardt2028} and is composed of a \SI{0.5}{\milli\meter} thick circular silicon wafer with a diameter of \SI{25}{\milli\metre}, coated with a \SI{100}{\nano\meter} layer of titanium acting as carrier substrate for the electro-deposited \textsuperscript{233}U. It has a central hole with a diameter of \SI{5}{\milli\metre} allowing for laser access along the ion axis. The source in its mount is shown in Fig.~\ref{fig:uraniumsource}. The mount is insulated from the carrier flange through the use of ceramic spacers and screws, thus allowing the source to be biased with an offset potential.

\begin{figure}[htb]
		\centering
		\includegraphics[width=\linewidth]{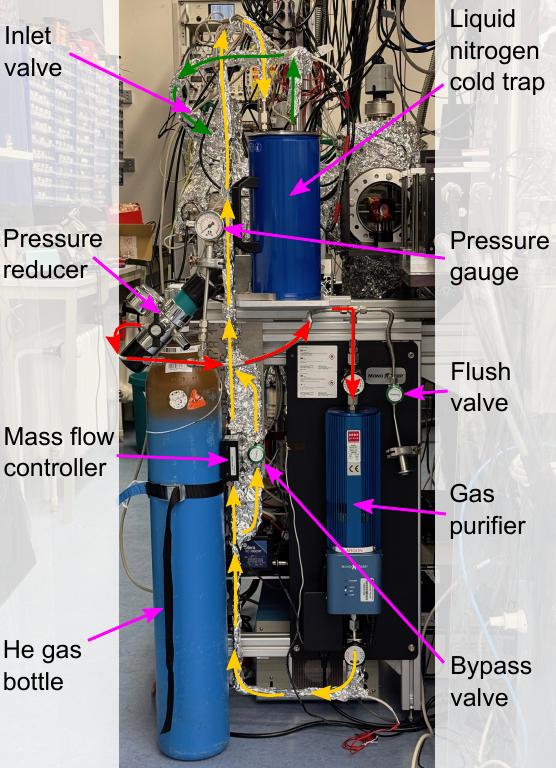}
		\caption{Photograph of the gas purification system.}
		\label{fig:gaspurifier}
\end{figure}

Since the produced \textsuperscript{229(m)}Th recoil nuclei leave the source with kinetic energies up to \SI{84}{\kilo\electronvolt} \cite{WakelingThesisWSU}, they have to be thermalized before extraction. This is achieved by collisional cooling in the ultra-pure helium gas inside the buffer-gas stopping cell. As impurities within the buffer gas cause charge exchange through collisions, thus lowering the ion extraction efficiency from the cell, the helium gas from a gas bottle (grade 6.0) is further purified. The gas purification system is shown in Fig.~\ref{fig:gaspurifier}. Using electropolished stainless-steel tubing with a diameter of 1/4 inch, the gas is first fed through a heated catalytic gas purifier (\textit{MonoTorr PS3-MT3}, SAES Group) and then through a liquid-nitrogen cold trap for the freeze out of any remaining contaminants. This cold trap consists of 19 windings of electropolished 1/4 inch tubing that are submerged in liquid nitrogen contained in a Dewar with a capacity of \SI{1.5}{\litre}. 
Additionally, in order to keep the buffer gas clean, the vacuum chamber is baked out at a temperature of \SI{130}{\degreeCelsius} for one day before operation. This assures background pressures below \SI{5e-9}{\milli\bar}. Further details are discussed in the context of the vacuum system in subsection~\ref{sec:VacuumSystem}. 
The mass flow of the helium gas into the buffer-gas cell is regulated by a mass flow controller (\textit{Aera FC780CHT}, Advanced Energy) installed after the purifier. This allows to define the buffer-gas pressure inside the cell, which is typically set to a value of \SI{32}{\milli\bar}. Under these conditions, the thorium ions will be stopped in the helium gas within a length of \SIrange[range-phrase=\,--\,,range-units=single]{1}{2}{\centi\meter} \cite{Thirolf19}, which allows for compact dimensions of the buffer-gas stopping cell.

\begin{figure}[htb]
    \centering
	\begin{subfigure}[c]{0.5\linewidth}
		\includegraphics [width=\linewidth]{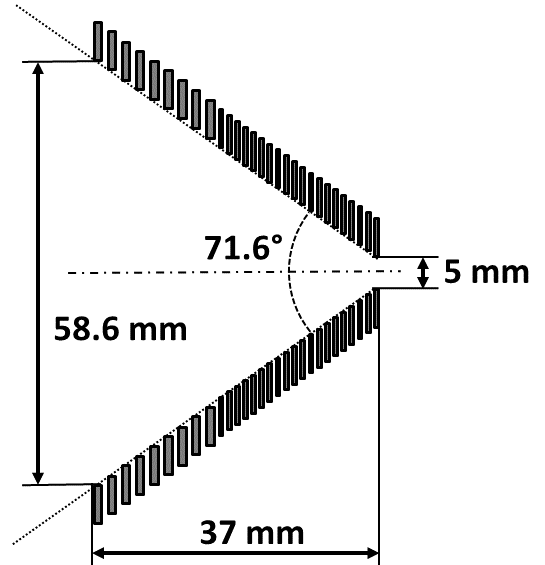}
	\end{subfigure}%
	\begin{subfigure}[c]{0.5\linewidth}
		\includegraphics[width=\linewidth]{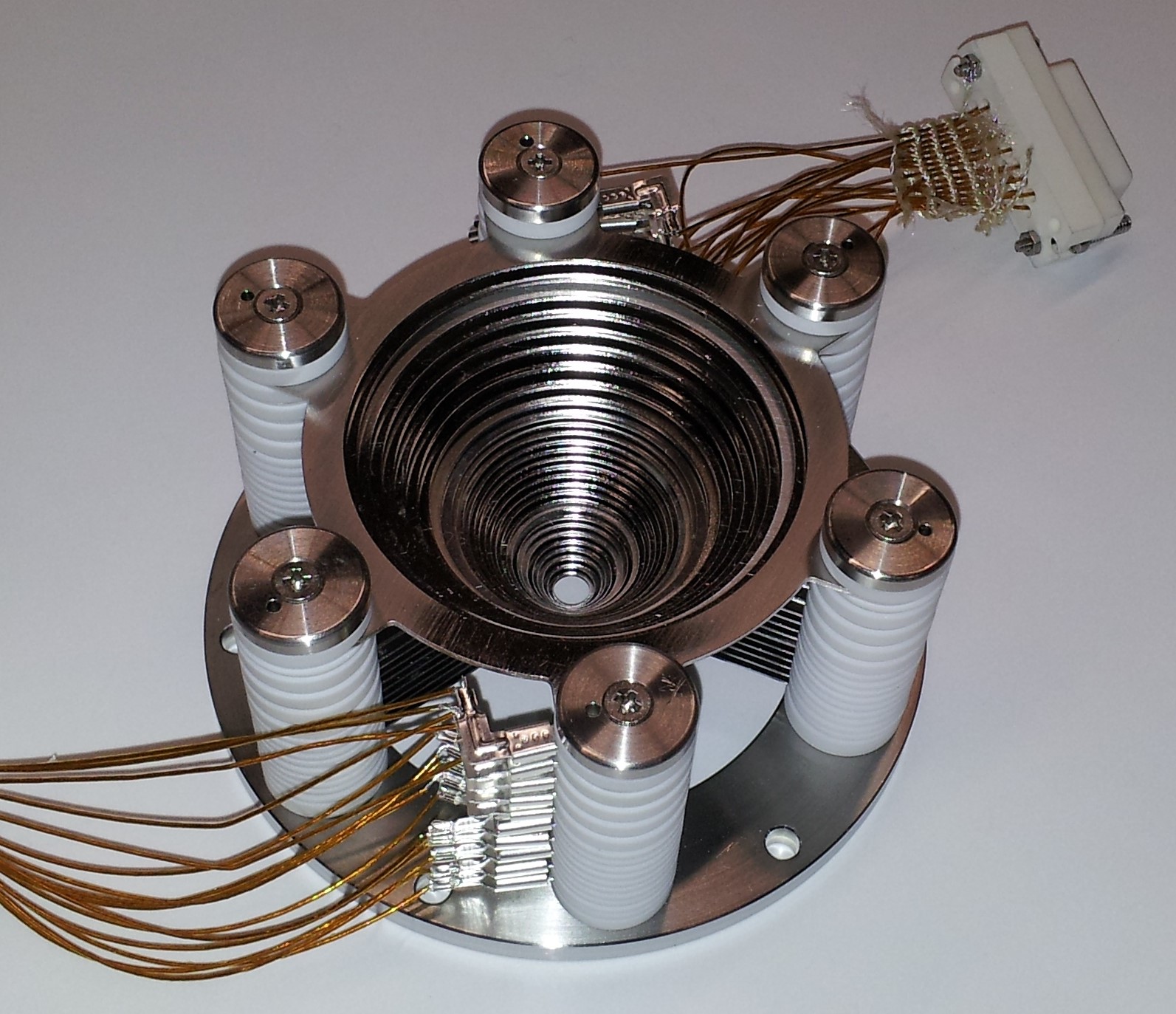}
	\end{subfigure}
    \caption{Left: schematic section view of the funnel electrodes. For the dimensions of the individual electrodes, see text. Right: photograph of the RF-DC funnel. Electrodes and ceramic spacers are stacked onto six mounting rods. The ring electrodes are electrically contacted via Kapton-insulated wires ending in UHV-compatible ceramic plug connectors.}
	\label{fig:funnelschematic}
\end{figure}

In order to extract the ions from the buffer gas, a conical RF-DC-funnel is used to guide the ions towards a nozzle separating the helium-filled chamber from the subsequent Extraction RFQ chamber. This ion funnel consists of 29 circular electrodes starting from an inner diameter of \SI{58.6}{\milli\meter} decreasing to an inner diameter of \SI{5.0}{\milli\meter} under an opening angle of \SI{71.6}{\degree}, as illustrated in Fig.~\ref{fig:funnelschematic}. The first 9 electrodes facing the source have a thickness of \SI{1}{\milli\meter} and are separated by a distance of \SI{1}{\milli\meter}. Their inner diameter decreases to \SI{34.4}{\milli\meter} in equidistant steps. In a distance of \SI{0.5}{\milli\meter} from the last of these electrodes, the 20 subsequent electrodes, each having a thickness of \SI{0.5}{\milli\meter}, are mounted with a reduced spacing of \SI{0.5}{\milli\meter}. Their inner diameter decreases from \SI{31.6}{\milli\meter} for the first of these electrodes to \SI{5.0}{\milli\meter} for the last electrode facing the de Laval nozzle. The overall length of the electrode structure thus is \SI{37.0}{\milli\meter}. The electrodes of the funnel are made of stainless-steel sheet metal and are stacked onto six ceramic rods, which provide alignment. The insulators creating the spacing between them are made of glass ceramics (\textit{Vitronit}, VITRON Spezialwerkstoffe). All electrodes have been electropolished to achieve a smooth surface in order to reduce the possibility of high-voltage sparking between neighbouring electrodes. The polarity of the RF voltage alternates between subsequent electrodes, which creates a repelling force for ions near the surface of the funnel. Typically, the RF-DC-funnel is operated at a resonance frequency of \SI{800}{\kilo\hertz} and amplitude of \SI{90}{\volt pp} with a superimposed DC voltage difference of \SI{15}{\volt} from the first funnel electrode to the last electrode that is closest to the nozzle, resulting in a gradient of $\approx \SI{4}{\volt\per\centi\meter}$ in axial direction.

The nozzle features a convergent-divergent de Laval geometry and is fabricated from stainless steel. The opening facing the funnel has a diameter of \SI{2.0}{\milli\meter} and tapers with an opening angle of \SI{90}{\degree} down to \SI{0.4}{\milli\meter} diameter at the throat of the nozzle. The length of the throat is \SI{0.3}{\milli\meter}. The diameter of the nozzle facing the subsequent chamber widens to \SI{6}{\milli\meter} at an opening angle of \SI{38}{\degree}. The nozzle is isolated from its carrier flange via a metal-ceramic soldering connection. This allows for the use of the nozzle as an electrode. It is typically set to a voltage about $\SI{0.7}{\volt}$ above the voltage at the last funnel electrode.
The geometry of the nozzle in combination with the difference in pressure between the buffer-gas cell at \SI{32}{\milli\bar} and the subsequent chamber, which is typically pumped to about \SI{e-3}{\milli\bar}, creates a super-sonic gas jet, dragging the ions off their guiding electric field lines and into the subsequent Extraction RFQ chamber.

The extraction efficiency of the buffer-gas stopping cell has been determined by a measurement for which the ion detection segment has been directly mounted after the Ion Guide segment, see Fig.~\ref{fig:setupoverview}. For typical operational parameters mentioned above, a count rate of approximately 200 \textsuperscript{229(m)}Th ions per second has been observed. 
Assuming a 100\,\% transmission efficiency through the Extraction RFQ and Ion Guide, as well as 100\,\% detection efficiency of the CEM, allows us to derive a lower limit to the extraction efficiency of the buffer-gas cell. We further assume that 50\,\% of the recoil ions from the source are emitted into the helium buffer gas. With these values, we arrive at an extraction efficiency of the buffer-gas cell of $>4$\,\%.

\begin{figure}[htb]
		\centering
		\includegraphics[width=\linewidth]{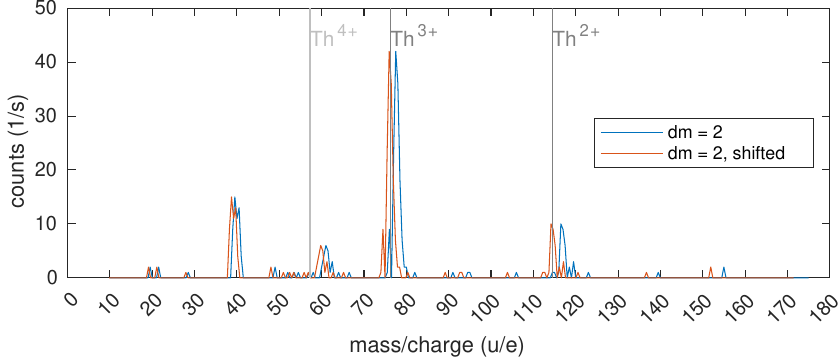}
		\caption{Mass spectrum of the ions extracted from the buffer-gas cell. Acquired utilizing the mass selectivity of the Ion Guide with the CEM detector mounted directly after the Ion Guide. The original spectrum recorded (blue) has been shifted (red) by the calibration factor of the RF amplitude.}
		\label{fig:Th_IG_massspectrum}
\end{figure}

A rough mass spectrum where the mass selectivity was provided by the Ion Guide is shown in Fig.~\ref{fig:Th_IG_massspectrum}. The resolution is limited by the voltage source of the Ion Guide to about \SI{2}{u/e} and a possible contamination with \textsuperscript{233}U and its other daughter products cannot be resolved. However, it shows that other contaminations are limited to one component at mass 39 and a small component at mass 60. Furthermore, it shows that only \textsuperscript{229(m)}Th in charge states of $3+$ or lower are extracted from the buffer-gas cell.

\subsection{The Extraction RFQ}
\begin{figure}[htb]
    \centering
    \includegraphics[width=\linewidth]{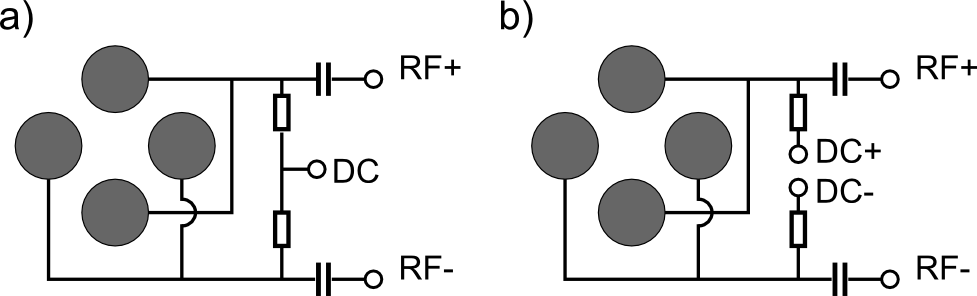}
    \caption{Schematic drawing indicating the way RF and DC voltages are applied to the Extraction RFQ and Paul trap (a) as well as the Ion Guide and QMS (b). Note that only the Ion Guide and the QMS are thus mass selective.}
    \label{fig:rf-dc-scheme}
\end{figure}

In the chamber subsequent to the buffer-gas cell, a segmented radio-frequency quadrupole (RFQ) is mounted on-axis with the de Laval nozzle at a distance of \SI{1}{\milli\meter} from the nozzle exit, see Fig.~\ref{fig:stoppcelloverview}. This Extraction RFQ with an overall length of \SI{174}{\milli\meter} consists of four rods segmented into eight electrodes, with the first electrode after the de Laval nozzle having a length of \SI{28}{\milli\meter}, followed by four electrodes with lengths of \SI{23}{\milli\meter} each, two shorter segments, each with a length of \SI{11.5}{\milli\meter}, and the final electrode with a length of \SI{24}{\milli\meter}. The cylindrical electrodes with an outer diameter of \SI{11}{\milli\meter} are made from stainless steel and are arranged with an ion-axis-to-electrode distance of \SI{4.8}{\milli\meter}. They are mounted on insulating Al\textsubscript{2}O\textsubscript{3} rods. A spacing of \SI{1}{\milli\meter} between the electrodes is ensured by ceramic insulators. All four electrodes of one segment are biased with the same DC voltage such that the RFQ is not mass selective. The segmentation of the Extraction RFQ, however, allows for applying a DC gradient along the beam axis to drag the ions from the nozzle all the way to the exit aperture of the Extraction RFQ chamber. The aperture has an inner diameter of \SI{2}{\milli\meter} and is the only opening in the gas-tight aluminum shield between the Extraction RFQ chamber and the adjoining Ion Guide, which allows for differential pumping.
As a result of the segmentation, the RFQ can be used as a buncher. For this purpose the sixth, seventh and the last segment (counted from the nozzle towards the Ion Guide) are equipped with electronics to allow for fast (few \si{\micro\second}) switching of applied DC voltages, assuring the rapid exit of the ions from the trapping region.
The eight segments are biased individually with DC voltages up to \SI{100}{\volt} (\textit{NHS 60 01p}, iseg), which are mixed with an RF voltage common to all segments. The circuit diagram of the RF-DC mixing for one segment is shown in Fig.~\ref{fig:rf-dc-scheme}~(a). The frequency of the RF voltage is set to \SI{790}{\kilo\hertz} and RF amplitudes up to \SI{360}{\volt pp} are applied.

\subsection{Ion Guide and Sr ablation source}
After the thorium Extraction RFQ, a separate vacuum chamber follows, which acts as an additional differential pumping stage before the QMS 2 and the ion trap. This additional differential pumping stage is necessary to reach the required vacuum for Coulomb crystallization in the ion trap. An aperture with a diameter of \SI{3}{\milli\meter} centred at the ion axis is the only opening to the adjoining QMS 2. It is also responsible for shielding the different RF fields from each other. A quadrupole Ion Guide with a length of \SI{269}{\milli\meter} and the same electrode dimensions as the previous Extraction RFQ is mounted in the chamber. It serves to guide the thorium ions from the Extraction RFQ to the subsequent QMS 2 chamber. The Ion Guide shares the RF voltage with the Extraction RFQ. The resulting resonant circuit has an eigenfrequency of \SI{790}{\kilo\hertz} and RF amplitudes up to \SI{360}{\volt pp} are applied. To be able to use the Ion Guide as a coarse mass filter, the applied RF voltages are mixed with DC voltages ranging between $\pm \SI{800}{\volt}$ (\textit{MHV-4}, Mesytec) according to Fig.~\ref{fig:rf-dc-scheme}~(b).
The amplitudes of the positive and negative DC voltages are controlled by a LabVIEW program on the laboratory PC.

The RF voltages applied to the electrodes of the Extraction RFQ as well as the Ion Guide are provided by an RF generator developed at the Justus-Liebig-Universität Gießen \cite{TimoDickel}, which allows for the amplitudes of the generated RF voltages $U_{RF}$ to be set by DC input voltages $U_{DC,in}$, while also providing DC output voltages $U_{diag}$ proportional to the RF amplitudes of its outputs. Therefore, $U_{diag}$ can be used as a process variable in a proportional-integral-derivative (PID) feedback loop that uses $U_{DC,in}$ as a control variable to actively stabilize the generated RF voltage amplitudes $U_{RF}$ and match the amplitudes of both phases with respect to each other.
This is implemented by digitizing the probe signals $U_{diag}$ of the two RF phases with an analog-to-digital converter card (\textit{NI-9215}, National Instruments) of an field-programmable-gate-array (FPGA) module (\textit{cRIO-9064}, National Instruments). PID regulators programmed by LabVIEW on the laboratory PC and running in real time on the cRIO-9064 FPGA module calculate control voltages that are sent from an analog-output card (\textit{NI-9264}, National Instruments) to the control input of two power supplies (\textit{Voltcraft PPS-16005}, Conrad Electronics). The outputs of these power supplies are then directly connected to the amplitude controls $U_{DC,in}$ of the RF voltage generator. Simultaneously scanning the amplitudes of the DC and RF voltages allows using the Ion Guide as a rough mass spectrometer, with the resolution limited to about \SI{2}{u/e} by the relatively large step size of the DC voltage source.

In addition, the Ion Guide is used as an ion catcher for laser-ablated \textsuperscript{88}Sr\textsuperscript{+} ions, which are required for the sympathetic laser cooling of \textsuperscript{229(m)}Th\textsuperscript{3+} in the Paul trap. The principle of the \textsuperscript{88}Sr\textsuperscript{+} ablation has already been described in a previous publication \cite{ScharlSetup2023}, however, there are some changes to the setup: The solid state ablation target made from SrTiO\textsubscript{3} (\textit{Strontium Titanate Single Crystal Substrate \textless100\textgreater, 634689-1EA}, Sigma Aldrich / Merck) is mounted at a distance of roughly \SI{16}{\milli\meter} from the ion axis, such that the frequency-doubled Nd:YAG ablation laser (\textit{Q-switched Nd:YAG laser, SN: \#03022502}, Quantel USA) can be aligned through a viewport on the opposite side of the vacuum chamber, see also Fig.~\ref{fig:setupoverview}. Laser pulses with a wavelength of \SI{532}{\nano\meter}, energies of typically \SI{2.5}{\milli\joule}, and durations of around \SI{5}{\nano\second} are focused to roughly \SI{300}{\micro\meter} onto the target surface. For each laser pulse, several 1000 ions are ablated and detected further downstream with the CEM detector.

The spatial separation of the ablation from the cryogenic trapping region in our setup is a result of the requirement to trap a hundred or more \textsuperscript{88}Sr\textsuperscript{+} ions without additional photoionization lasers while also assuring that the risk of coating of the trap electrodes due to the hardly quantitatively controllable output of neutral and charged particles in the ablation process is minimized. Such a coating would generate patch potentials on the trap electrodes, which in turn would negatively influence the motional heating rate or position of the trapped in ions \cite{turchette2000heating, deVoe2002experimental, daniilidis2011fabrication, haerter2014long}.

\begin{figure*}[t]
	\centering
	\includegraphics[width = 1\textwidth]{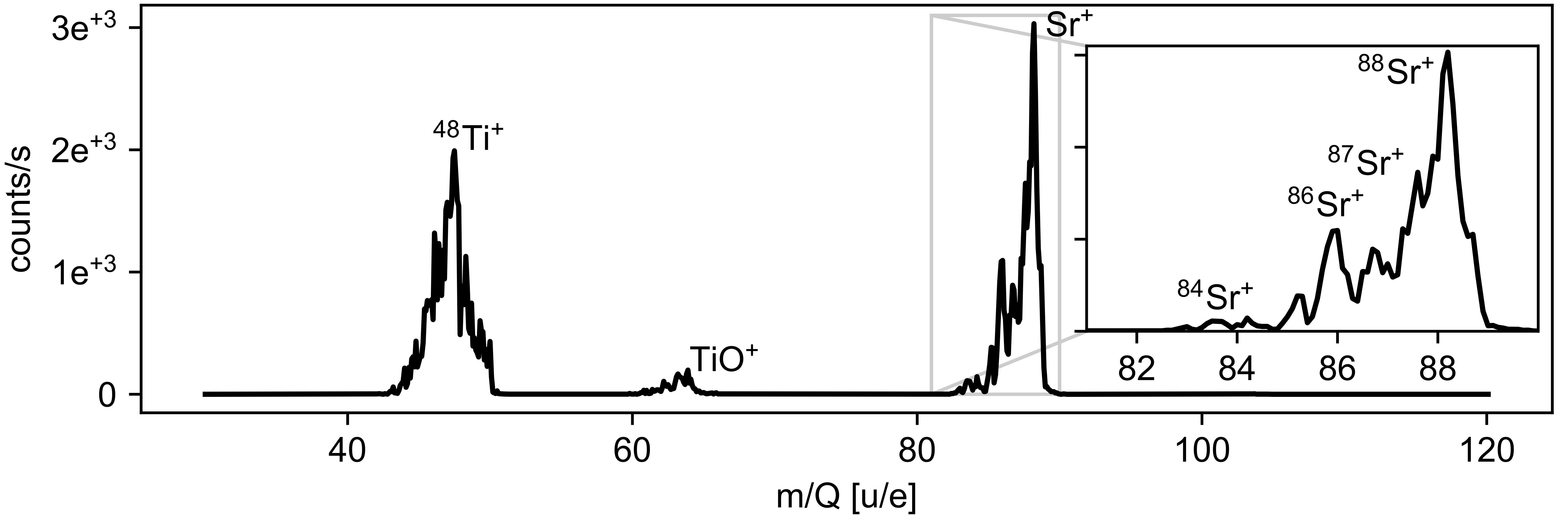}
	\caption{Mass scan of ablated ions from the SrTiO\textsubscript{3} target in the Ion Guide performed with QMS 2. The filter width is $\Delta(m/Q)=\SI{2}{u/e}$ and the integration time is \SI{1}{\second} per scan step of \SI{0.1}{u/e}. The ablation laser is operated with a repetition rate of \SI{5}{Hz}.
    \label{fig: Sr MassScan}}
\end{figure*}

Thorium ions from the Extraction RFQ as well as the ablated strontium ions are not yet isotopically pure at this stage of the beam line. Therefore, the ions need to undergo a mass filtering process before reaching the trap. A full mass scan of the laser-ablated ions performed with the QMS 2 quadrupole mass separator (described in the following subsection) is shown in Fig.~\ref{fig: Sr MassScan}. The mass spectrum shows the typical composition of the SrTiO\textsubscript{3} target with the natural abundance of strontium isotopes being resolved.

\subsection{The Quadrupole Mass Separators}
\label{section:QMSexp}
The two quadrupole mass separators located on each side of the central cryogenic Paul trap are identical, and both are described in this subsection. They serve two purposes: Firstly, the QMS located between the buffer-gas stopping cell and the Paul trap (QMS 2) is used to select the species and charge state of the ions loaded into the trap. Therefore, it has to separate the \textsuperscript{229}Th\textsuperscript{3+} ions from other accompanying daughter products of the $\alpha$-decay chain of \textsuperscript{233}U also contained in the ion beam extracted from the buffer-gas stopping cell. Furthermore, the ablated \textsuperscript{88}Sr\textsuperscript{+} ions need to be separated from titanium, oxygen, and other isotopes of strontium produced in fractions according to natural abundance. For this purpose, a relative resolution of about $(m/Q)/\Delta (m/Q)=150$ is targeted. This resolution is sufficient to reliably separate \textsuperscript{229}Th from \textsuperscript{225}Ra, which is the closest in mass of the daughter products from the \textsuperscript{233}U decay chain, as well as from \textsuperscript{233}U, which is also released from the source. In the $3{+}$ charge state, these two unwanted ion species are separated from \textsuperscript{229}Th\textsuperscript{3+} by $\Delta(m/Q)=1.33$, resulting in a minimum required resolution of $(m/Q)/{\Delta (m/Q)}=57.25$. The most strict requirement is set by the separation of \textsuperscript{88}Sr\textsuperscript{+} from \textsuperscript{87}Sr\textsuperscript{+} resulting in a required minimum resolution of $(m/Q)/{\Delta (m/Q)}=88$, while all other requirements set by charge state selection or separating \textsuperscript{88}Sr\textsuperscript{+} from titanium are less strict. It should be noted that the discussed requirements omit the trace contamination of the \textsuperscript{233}U source with \textsuperscript{232}U, which originates from the production process of the source \cite{Wense15}. While a resolution of $(m/Q)/{\Delta (m/Q)}=150$ will be sufficient to separate \textsuperscript{229}Th from \textsuperscript{232}U in the $3{+}$, $2{+}$ and $1{+}$ charge state, the separation of \textsuperscript{229}Th\textsuperscript{3+} from \textsuperscript{228}Th\textsuperscript{3+} would require a significantly higher resolution due their small separation by $\Delta(m/Q)=0.33$. As the contribution of \textsuperscript{228}Th to the amount of extracted ions can be estimated as ${N_{228\text{Th}}}/{N_{229\text{Th}}}\approx\num{9e-4}$
\footnote{Assuming a contamination of the \textsuperscript{233}U source of ${N_{232\text{U}}}/{N_{233\text{U}}}=\num{3.9e-7}$ based on previous work with identical source material \cite{Wense15}, the ratio of ${N_{228\text{Th}}}/{N_{229\text{Th}}}$ can be estimated by  ${N_{228\text{Th}}}/{N_{229\text{Th}}}=(\lambda_{232\text{U}}\cdot N_{232\text{U}})/(\lambda_{233\text{U}}\cdot N_{233\text{U}})$. With $\lambda={\ln{2}}/{T_{1/2}}$ and the half-lives of \textsuperscript{233}U and \textsuperscript{232}U being \SI{1.592e5}{years} and \SI{68.9}{years}, respectively \cite{NNDCnudat}, one finds ${N_{228\text{Th}}}/{N_{229\text{Th}}}=\num{2.31e3} \times \num{3.9e-7}= \num{9.0e-4}$}
it is deemed negligible.

Secondly, the quadrupole mass separator between the Paul trap and the ion detector (QMS 1) serves as an ion guide towards the detector while enabling the investigation of possible charge state changes after prolonged trapping by providing charge state selectivity.

\subsubsection{Electrode geometry}
The design properties of both quadrupole mass separators match the ones proposed in \cite{HaettnerThesis,Haettner2018}. They consist of four rods with a radius of $r=\SI{9}{\milli\meter}$ made from stainless steel that are segmented into three electrodes each, accounting for an overall length of \SI{402}{\milli\meter}. While the region of mass selectivity is created by the central segment with a length of \SI{300}{\milli\meter}, the two other segments with a length of \SI{50}{\milli\meter} on each end of the rods serve as Brubaker lenses \cite{Brubaker1968}. Only the RF voltages, but not the DC voltages of the mass selective segments are applied to the segments of the Brubaker lenses. This configuration reduces the defocusing by fringe fields and thus significantly improves the overall ion transmission efficiency at a given mass resolution.
The spacing between the electrodes amounts to \SI{1}{\milli\meter} and is provided by ceramic spacers. With an ion-axis-to-electrode distance of $r_{0}=\SI{7.98}{\milli\meter}$, a ratio of ${r}/{r_{0}}=1.128$ is achieved, being close to the optimal design value of 1.13 found by Douglas and Konenkov \cite{Douglas2002}. The electrodes are surrounded by a grounded cylindrical stainless steel shielding of \SI{1}{\milli\meter} thickness with an inner radius of $r=\SI{32}{\milli\meter}$ forming a perforated cylinder with an open area of \SI{51}{\percent} in order to achieve good vacuum conditions inside the device. Both quadrupole mass separators are equipped with apertures facing the trap, which have a diameter of \SI{5}{\milli\meter}, providing the shielding between the field regions of the QMS and the linear Paul trap, allowing for a smooth transition of the ions from one field region to the other.

\subsubsection{Signal processing electronics}
\label{sectionPID}
The operation of the quadrupole mass separators requires the application of RF as well as DC voltages to the quadrupole rods, see Fig.~\ref{fig:rf-dc-scheme}~(b). Similar to the RF of the Extraction RFQ and the Ion Guide, the RF voltage is provided by an RF generator developed at the Justus-Liebig-Universität Gießen \cite{TimoDickel}. Also these RF amplitudes are controlled in real time by the cRIO-9064 FPGA module steered by the laboratory PC.
In order to efficiently separate the $\alpha$ decay daughter products of \textsuperscript{233}U from the \textsuperscript{229}Th ions, the quadrupole mass separators are operated at a mass resolution of $R=(m/Q)/{\Delta (m/Q)}\approx150$. The achievable resolution of the QMS is directly affected by the precision of the applied voltages, described by the relation:
\begin{equation}
	\frac{\Delta U_{RF}}{U_{RF}} = \frac{\Delta U_{DC}}{U_{DC}} \leq \frac{1}{2R}.
	\label{eq:precisionQMS}
\end{equation} 
Therefore, a relative precision better than \SI{3.33e-3}{\volt} is required in order to reach the targeted resolution. In the case of the DC voltage, this requirement is met by choosing a sufficiently precise power supply (\textit{Traco THV 12-300P/N}, TRACO Power Group), which provides a nominal stability of \num{5e-4} \cite{TracoTHV}. To achieve the required precision for the RF amplitude, it is actively regulated with a PID control directly implemented on the cRIO-9064 FPGA module.

\begin{figure}[htb]
\centering
\begin{subfigure}[b]{\linewidth}
   \includegraphics[width=\linewidth]{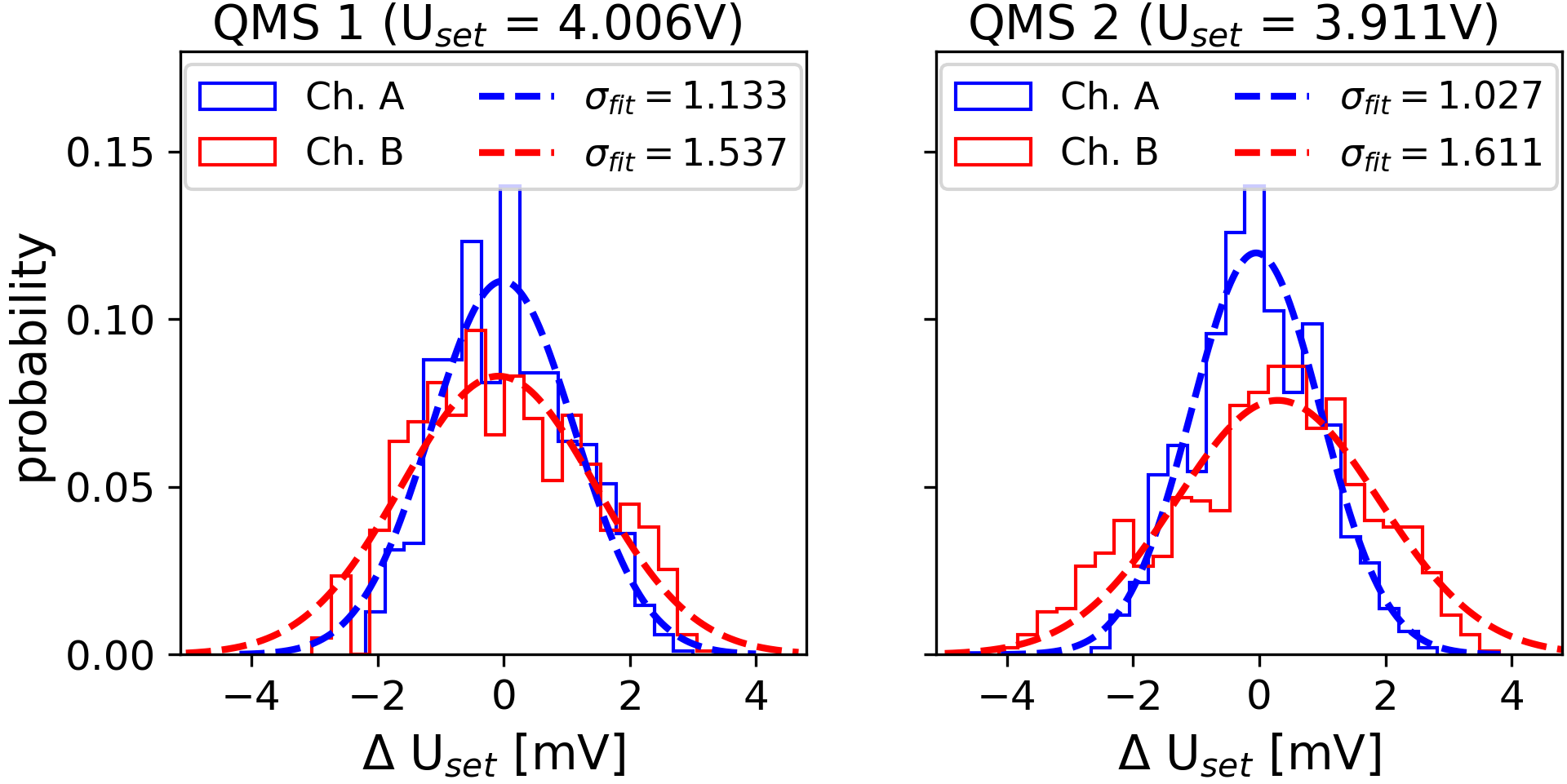} 
\end{subfigure}
\begin{subfigure}[b]{\linewidth}
   \includegraphics[width=\linewidth]{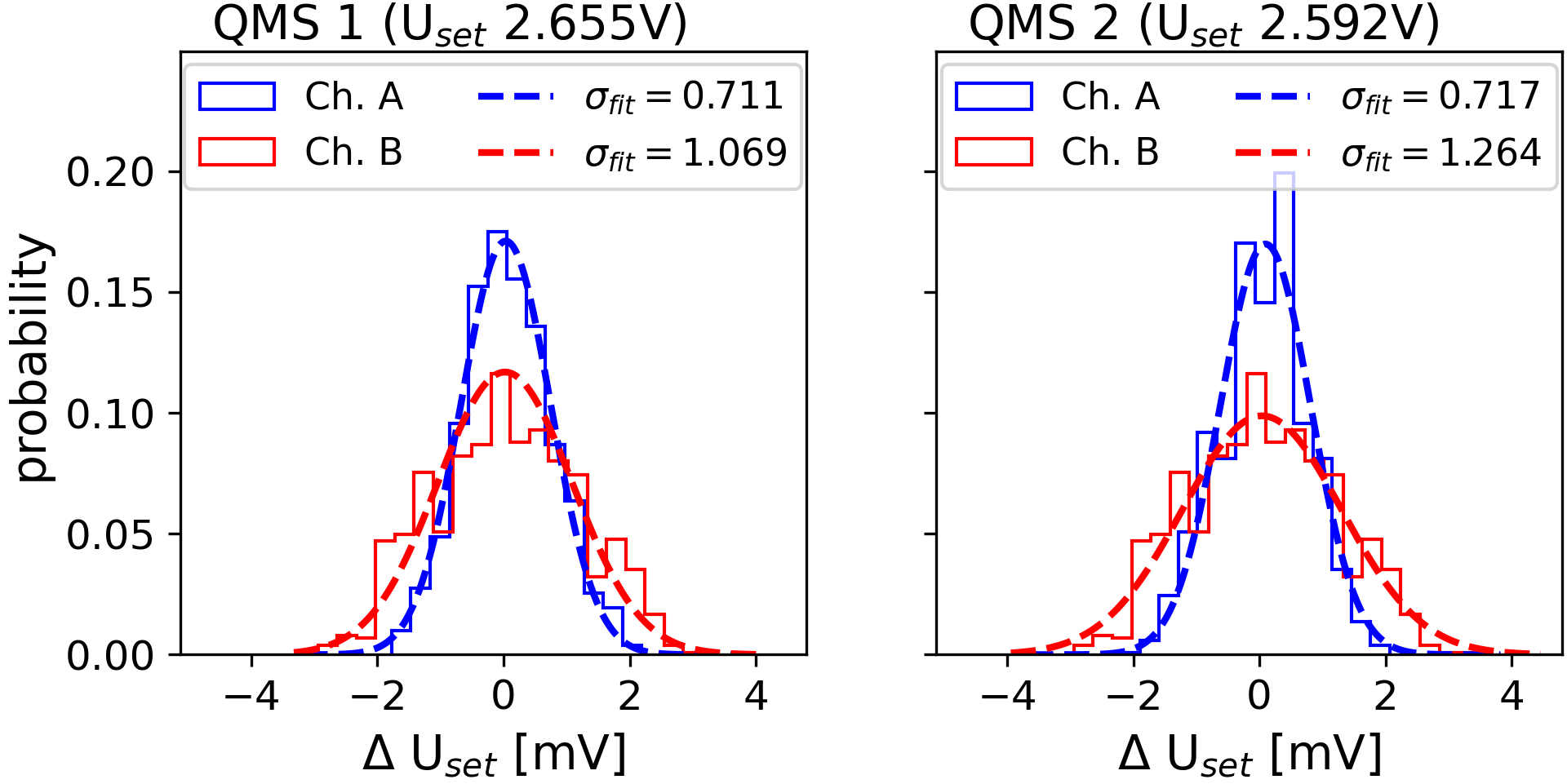}
\end{subfigure}
\caption{Fluctuations of the DC control voltages $U_{diag}$, being proportional to the amplitudes of the generated RF voltages, around their set-points corresponding to \textsuperscript{229}Th\textsuperscript{2+} (top panels) and \textsuperscript{229}Th\textsuperscript{3+} (bottom panels), for both quadrupole mass separators. Note the difference in the set points between the two QMS, originating from the slightly different resonance frequencies of both QMS modules.}
\label{fig:controllprecision}
\end{figure}

As the PID algorithm on the cRIO-9064 FPGA module processes digital signals, the diagnostic voltages $U_{diag}$ are digitized with the use of a second analog-to-digital converter (\textit{NI-9215}, National Instruments) in order to calculate a control variable in the PID feedback loop. Similarly, the digital output of the PID control loop is converted to an analog signal through the NI-9264 analog-output module. As the output of this module is limited to a nominal voltage of $\pm \SI{10}{\volt}$ and a maximum current of $\pm \SI{16}{\milli\ampere}$ \cite{NI9264}, the voltage provided by this module is not sufficient to serve as input voltage $U_{DC,in}$ for the RF generator, which requires a current of \SI{1}{\ampere}. Therefore, $U_{DC,in}$ is provided through a remotely controllable power supply (\textit{Voltcraft VSP2405}, Conrad Electronics) with the amplitude set by the control voltage provided by the NI-9264 module. By investigating the voltage variations in the RF amplitudes corresponding to \textsuperscript{229}Th\textsuperscript{2+} and \textsuperscript{229}Th\textsuperscript{3+} ions, respectively, it could be demonstrated that the required precision of better than \num{3.33e-3} can be reached. This becomes evident from the results obtained for both quadrupole mass separators listed in Tab.~\ref{TablePrecision}. The listed values are derived from the measured data shown in Fig.~\ref{fig:controllprecision}, by applying a Gaussian fit to the data. The bin width of \SI{0.305}{\milli\volt} of the data corresponds to the nominal resolution of the analog-to-digital converter NI-9215 (16 bit, $\pm\SI{10}{\volt}$) used to read out the control voltage. 

\begin{table}[htb]
	\centering
	\begin{tabular}{ccc}
		\hline\noalign{\smallskip}
		& $\frac{\Delta U}{U}$ (Channel A) & $\frac{\Delta U}{U}$ (Channel B)\\ 
		\noalign{\smallskip}\hline\noalign{\smallskip}
		QMS1 (\textsuperscript{229}Th\textsuperscript{2+}) & $\approx 6.66\cdot10^{-4}$ & $\approx 9.03\cdot10^{-4}$\\ 
		\\
		QMS1 (\textsuperscript{229}Th\textsuperscript{3+}) & $\approx 6.31\cdot10^{-4}$ & $\approx 9.48\cdot10^{-4}$\\ 
		\\
		QMS2 (\textsuperscript{229}Th\textsuperscript{2+}) & $\approx 6.18\cdot10^{-4}$ & $\approx 9.70\cdot10^{-4}$\\
		\\
		QMS2 (\textsuperscript{229}Th\textsuperscript{3+}) & $\approx 6.51\cdot10^{-4}$ & $\approx 1.15\cdot10^{-3}$ \\
		\noalign{\smallskip}\hline
	\end{tabular} 
	\caption{Relative control voltage ($U_{diag}$) uncertainties for both pairs of QMS rods (denoted as channel A and B) of both quadrupole mass separators (QMS 1, QMS 2) for the case where the mass filter is set to \textsuperscript{229}Th\textsuperscript{2+} and \textsuperscript{229}Th\textsuperscript{3+}, respectively.}
	\label{TablePrecision}
\end{table}

\subsubsection{QMS performance}
In a first step, the functionality of the quadrupole mass separators was investigated with the help of a calibration ion source. For this purpose, a heated alkali ion source (\textit{TB-118
Aluminosilicate Ion Sources}, Heatwave Labs) containing potassium, rubidium and caesium in natural isotopic composition was used. In contrast to the \textsuperscript{88}Sr\textsuperscript{+} ablation source, such a heated ion source delivers a continuous and stable ion beam, which is more favourable for calibration.
Furthermore, for these characterization measurements the segments from the buffer-gas cell to the Ion Guide were removed and the alkali source was directly mounted in a chamber before the QMS 2 chamber, compare Fig.~\ref{fig:setupoverview}.

\begin{figure}[htb]
	\centering
	\includegraphics[width=\linewidth]{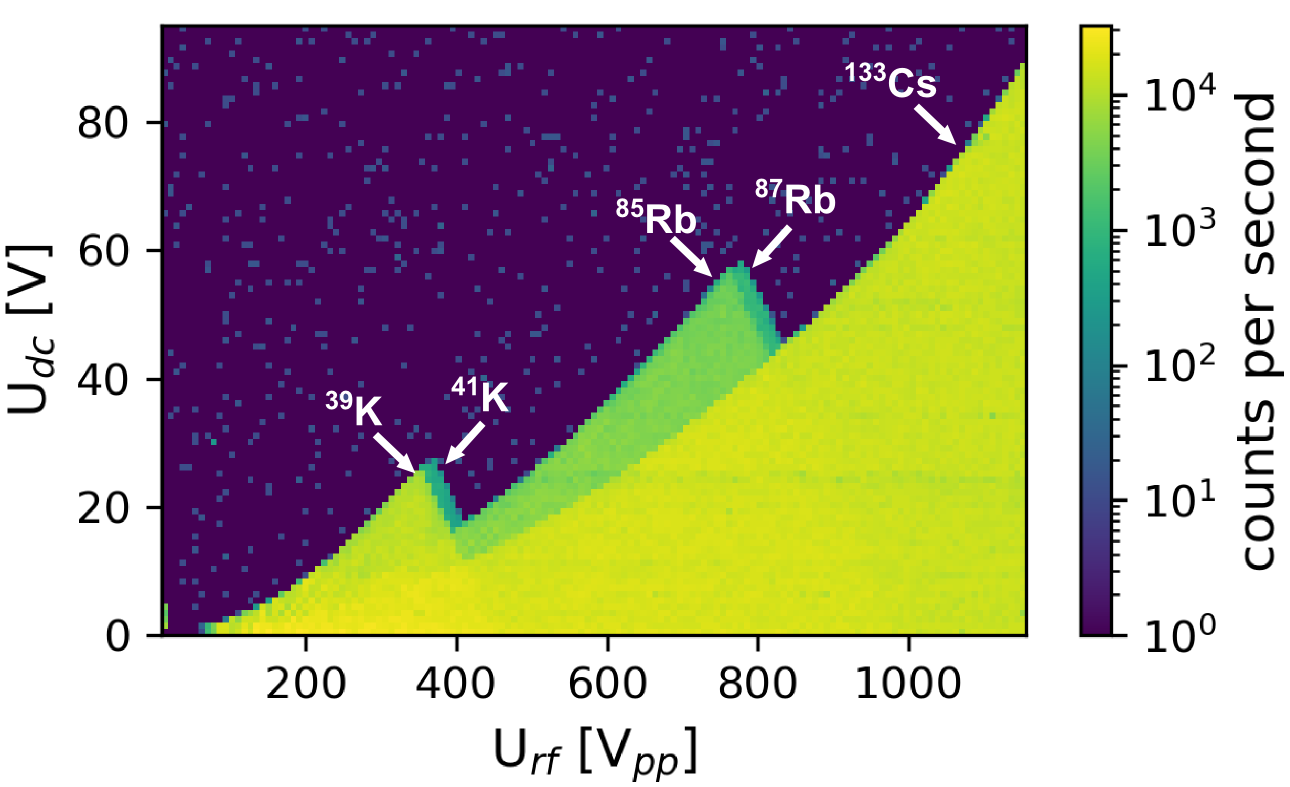}
	\caption{Mapping of the stability diagram of the quadrupole mass separator QMS 2 for the three elemental species contained in the calibration ion source.}
	\label{fig:stabilitydiagram}
\end{figure}

\begin{figure}[htb]
	\centering
	\includegraphics[width=\linewidth]{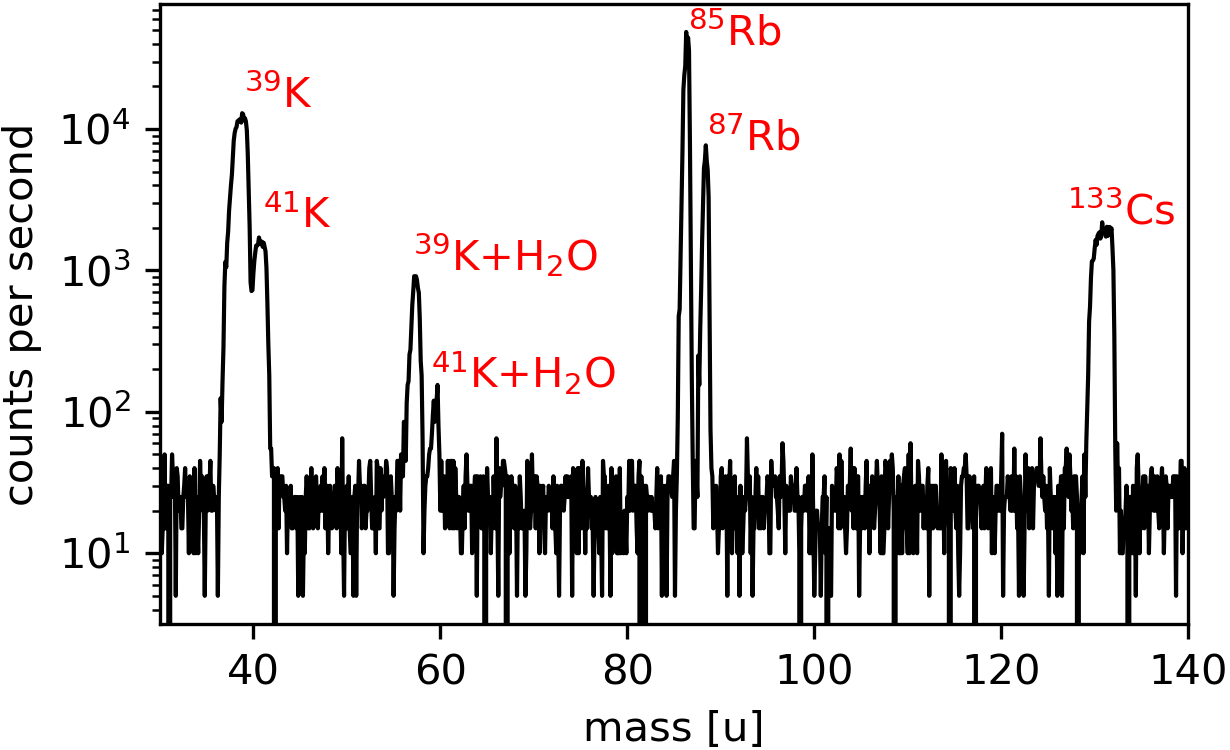}
	\caption{Mass spectra obtained with the calibration ion source. The observable contamination with water adducts can be traced back to an insufficient bake out.}
		\label{fig:massspectra}
\end{figure}

By scanning both the DC voltage in the range of \SIrange[range-phrase=\,--\,,range-units=single]{0}{95}{\volt} and the RF voltage in the range of \SIrange[range-phrase=\,--\,,range-units=single]{0}{1150}{\volt pp}, a stability diagram of QMS 2 was obtained. The result is shown in Fig.~\ref{fig:stabilitydiagram}, where the characteristic triangle-shaped region of stability for each of the ion species is labelled accordingly. The corresponding mass spectrum of the calibration ion source is plotted in Fig.~\ref{fig:massspectra}. The spectrum was obtained with QMS 1 by scanning  the RF and DC voltages simultaneously while keeping the ratio ${U_{RF}}/{U_{DC}}$ constant along the scan line. This leads to a varying peak width within a broad spectral range, as shown in Fig.~\ref{fig:massspectra}. Since both QMS modules are operated as static mass filters, either filtering for \textsuperscript{229}Th\textsuperscript{2+}, \textsuperscript{229}Th\textsuperscript{3+} or \textsuperscript{88}Sr\textsuperscript{+}, they have to be separately calibrated for the respective ion species in order to account for the varying peak width. From the results it becomes evident that the mass separators can be successfully operated in the voltage regions required to be selective for either \textsuperscript{88}Sr\textsuperscript{+}, \textsuperscript{229}Th\textsuperscript{2+} or \textsuperscript{229}Th\textsuperscript{3+} ions.

Furthermore, the transmission of the quadrupole mass separators at the targeted resolution of $(m/Q)/{\Delta (m/Q)} = 150$ was investigated and found to be $(30\pm10)\,\%$ if $\Delta (m/Q)$ is defined as the width of the mass peak at $10\,\%$ of its maximum height and $(67\pm10)\,\%$ when $\Delta (m/Q)$ is defined as the full width at half maximum (FWHM), respectively.

\subsection{The cryogenic linear Paul trap}
The cryogenic Paul trap was designed to provide long enough storage durations for \textsuperscript{229m}Th\textsuperscript{q+} ions ($q = 1,2,3$) to allow measuring the the thorium isomer's radiative lifetime in vacuum.
The radiative lifetime in vacuum has been theoretically predicted \cite{PhyRevC15Tkalya} and recently indirectly experimentally measured \cite{PhysRevTiedauSchaden24,elwellPhysRevLett24,NatureYamaguchi24} to be in the range of \SI{e3}{\second}. In order to reach these long storage times, the linear Paul trap is operated at cryogenic temperatures of about \SI{8}{\kelvin}
and the trapped \textsuperscript{229(m)}Th\textsuperscript{3+} ions are sympathetically laser cooled using \textsuperscript{88}Sr\textsuperscript{+}. The design of the cryogenic setup is based on the experience of the MPI Heidelberg group for highly charged ion dynamics gained with the CryPTEx cryogenic Paul trap \cite{schwarz2012cryogenic, leopold2019cryogenic}.

\subsubsection{The cryogenic environment}
\begin{figure}[htb]
	\centering
	\includegraphics[width=\linewidth]{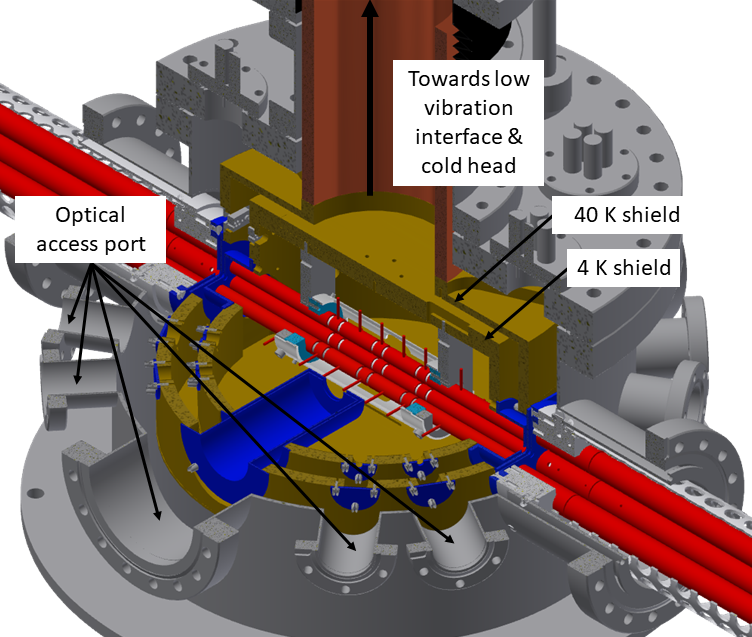}
	\caption{Section view of the cryogenic Paul trap setup, showing the linear Paul trap inside the nested structure of its surrounding heat shields. For better visibility, especially in the case of the pocket-shaped apertures for the fluorescence collecting lens system, the apertures are coloured in blue, while the two heat shields are depicted in gold.}
	\label{fig:schematicpaultrap}
\end{figure}

The chosen experimental goals define the following requirements for the cryogenic environment:
\begin{itemize}
	\item Reaching temperatures as low as possible in the trap region in order to achieve optimum vacuum conditions through condensation of the residual gas on the cold surfaces to reach long storage times;
	\item On-axis accessibility of the trap (for ion injection) and access perpendicular to the trap axis (for laser cooling and optical diagnostics), while minimizing the required open areas in the heat shielding to prevent thermal radiation from entering the trap;
	\item Suppressing vibrations generated by the cryocooler through vibrational decoupling from the trap mounting.
\end{itemize}
To reach the desired cryogenic temperatures at the trap, it is placed inside two nested cylindrical heat shields as seen in Fig.~\ref{fig:schematicpaultrap}. The inner heat shield is kept at a temperature of about \SI{8}{\kelvin}, while the outer heat shield is kept at a temperature of about \SI{42}{\kelvin}, providing an intermediate temperature stage between the \SI{8}{\kelvin} stage of the trap and the room temperature of the vacuum chamber. The temperatures are measured and monitored using silicon diode temperature sensors (\textit{DT-670B-CU}, Lake Shore Cryotronics) read out by the corresponding Lake Shore temperature monitor (\textit{Model 218}, Lake Shore Cryotronics). The heat shields are fabricated from oxygen-free high conductivity (OFHC) copper. In order to prevent the absorption of thermal radiation, the shields are coated with a \SI{0.5}{\micro\meter} gold layer applied on top of a \SIrange[range-phrase=\,--\,,range-units=single]{8}{10}{\micro\meter} silver layer acting as a diffusion barrier between the gold layer and the copper. Furthermore, the gold coating prevents oxidization of the copper when the system is vented, thus maintaining high reflectivity for thermal radiation.

The inner diameters of the inner and outer heat shields are \SI{220}{\milli\meter} and \SI{296}{\milli\meter}, respectively. Their wall thickness of \SI{10}{\milli\meter} was chosen in order to also effectively shield external electric fields. Furthermore, both heat shields are equipped with bottom plates of \SI{2}{\milli\meter} thickness. The cooling is provided by a two-stage pulse tube cryocooler (\textit{RP-082B2-F70H}, Sumitomo Heavy Industries) with a nominal cooling capacity of \SI{40}{\watt} at \SI{45}{\kelvin} in the first stage and \SI{1}{\watt} at \SI{4.2}{\kelvin} in the second stage \cite{Coolinghead}. 

\begin{figure}[htb]
	\centering
	\includegraphics[width=\linewidth]{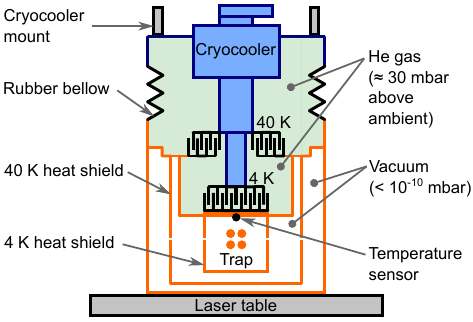}
	\caption{Sketch of the low-vibration cryocooler interface \cite{ColdEdgeVibrationInterface}. The sections in orange and blue are only connected by a rubber bellow.}
	\label{fig:lowvibrationinterface}
\end{figure}

In order to suppress the vibrations generated by the cold head, the cold head can be mounted to the ceiling of the laboratory and connected to the chamber through an ultra-low-vibration (ULV) interface from ColdEdge Technologies \cite{ColdEdgeVibrationInterface}. 
The ULV interface consists of a tube filled with helium (grade 5.0) acting as an exchange gas to transfer the cooling power of the cold head to the Paul trap, see Fig.~\ref{fig:lowvibrationinterface}. The thermal connection is assured by the helium gas between the interleaved copper fingers. If mounted to the ceiling, the only mechanical connection between the cooling head and the trap chamber is a rubber bellow required for gas tightness. The lower part of the ULV interface is directly mounted to the vacuum chamber of the trap and provides two vibrationally decoupled stages that correspond to the two stages of the cryocooler. The heat shields are connected to these two stages. For more information on the working principle and design of the ULV interface we refer to \cite{dubielzig2021ultra, Dubielzig_Thesis}.

\begin{figure}[htb]
	\centering
	\includegraphics[width=\linewidth]{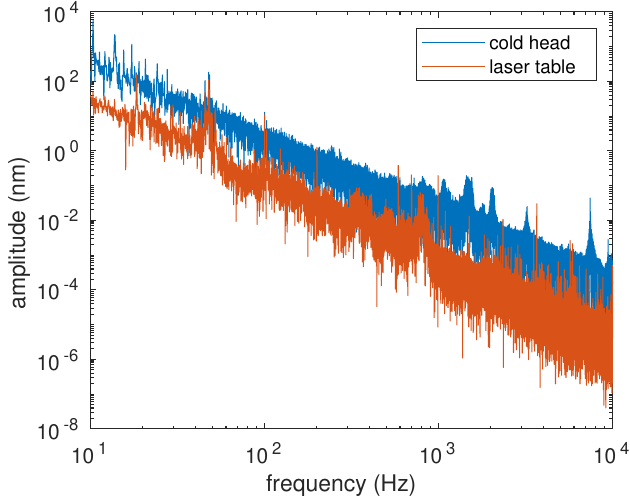}
	\caption{Vibration amplitudes at the cold head and laser table.}
	\label{fig:vibration_amplitudes}
\end{figure}

The vibration-isolation properties of the ULV interface were tested with a piezoelectric vibration sensor (\textit{PCB-352C33}, PCB Piezotronics) read out with a signal conditioner (\textit{PCB-682A02}, PCB Piezotronics) and an oscilloscope. For these measurements, the sensor was mounted directly on the cryocooler cold head (near the cryocooler mount) for the characterization of cold head vibrations and on the laser table for characterization of the vibrations transmitted to the laser table. The measured vibration amplitudes at both locations are compared in Fig.~\ref{fig:vibration_amplitudes}. The vibrations at the laser table are attenuated by about 2 orders of magnitude, and the residual vibration amplitude is below \SI{10}{\nano\meter} for frequencies above \SI{60}{\hertz}. A more elaborate characterization of the system was carried out in \cite{dubielzig2021ultra}, and the resulting vibration amplitudes should also be applicable to our system.

\subsubsection{Optical access}
The trap region is accessible for lasers by incoupling along the ion axis through CF40 viewports, one in the buffer-gas stopping cell behind the uranium source and one in the ion detection chamber. To provide free passage for the light, the ion detector can be moved perpendicular to the ion axis (see also subsection~\ref{sec: Ion Detection} and Fig.~\ref{fig:setupoverview}). For further laser manipulation and fluorescence diagnostics of the trapped ions, the heat shields contain eight additional openings with a \SI{10}{\milli\meter} diameter in the inner shield and \SI{16.5}{\milli\meter} in the outer shield, respectively, that allow for placing apertures right in front of the corresponding openings. These eight openings provide four lines of sight with angles of \SI{30}{\degree}, \SI{50}{\degree}, \SI{130}{\degree}, and \SI{150}{\degree} with respect to the trap axis for laser access. In order to limit the amount of thermal radiation reaching the trap, all openings are equipped with apertures reducing the opening to a diameter of \SI{5}{\milli\meter}. These \SI{2}{\milli\meter} thick apertures are made from OFHC copper and are coated with gold in the same way as the heat shields.

Furthermore, the shields contain two openings with a line of sight perpendicular to the trap axis with diameters of \SI{53}{\milli\meter} in the inner shield and \SI{70}{\milli\meter} in the outer shield, respectively. One of these openings is equipped with a lens system (see \cite{ScharlSetup2023}) collecting the fluorescence light of the trapped ions. In order to be positioned as close as possible to the trap, while simultaneously limiting the open area in the heat shields, this opening is equipped with pocket-shaped apertures reaching inside the heat shields as seen in Fig.~\ref{fig:schematicpaultrap}. These \SI{1}{\milli\meter} thick apertures are made from stainless steel and have an opening of \SI{20}{\milli\meter}. The length of the aperture at the outer heat shield is \SI{105}{\milli\meter}, and the length of the aperture at the inner heat shield is \SI{78}{\milli\meter}. Currently, the opposite openings in the heat shields are covered with gold-coated copper plates. In the future, these openings will be modified to allow for focusing laser light from a vacuum ultraviolet (VUV) frequency comb \cite{Wissenberg2025} onto a singe \textsuperscript{229}Th\textsuperscript{3+} ion.

\subsubsection{Trap assembly}
\begin{figure}[htb]
	\centering
	\includegraphics[width=\linewidth]{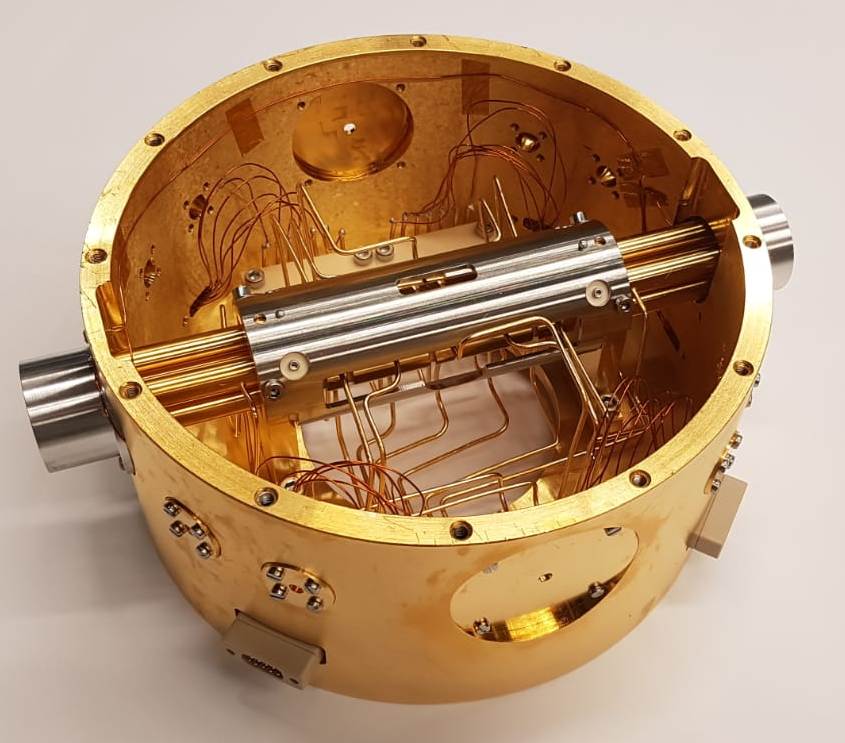}
	\caption{Photograph of the linear Paul trap mounted inside the gold-coated \SI{4}{\kelvin} heat shield with its access ports covered by the corresponding apertures. Note the openings in the stainless steel cylinder as well as the pocket-shaped apertures at both ends of the trap.}
	\label{fig:paultrapfoto}
\end{figure}

The Paul trap was designed to have a large trapping volume for the confinement of a large number of \textsuperscript{229(m)}Th\textsuperscript{3+} ions. It consists of four segmented rods with an overall length of \SI{282}{\milli\meter}. Each of the segmented rods consists of a central trapping electrode with a length of \SI{8}{\milli\meter} followed by an electrode of \SI{30}{\milli\meter} length separating it from another trapping electrode of \SI{8}{\milli\meter} length on each side of the center and ending with \SI{93}{\milli\meter} long electrodes at both sides. The electrodes have an outer diameter of \SI{11}{\milli\meter} and are arranged with an ion-axis to electrode distance of \SI{4.8}{\milli\meter}. The electrodes are fabricated from OFHC copper and separated by ceramic spacers made from Al\textsubscript{2}O\textsubscript{3}, resulting in a spacing of \SI{2}{\milli\meter} between them. To improve the electrical conductance of the electrode surfaces, they were coated with a \SI{0.5}{\micro\meter} layer of gold applied on top of an \SIrange[range-phrase=\,--\,,range-units=single]{8}{10}{\micro\meter} silver layer acting as a diffusion barrier between the gold layer and the copper. The electrodes and ceramic spacers are threaded on an insulating Al\textsubscript{2}O\textsubscript{3} rod and held in place by being pressed together by the terminating electrodes at both ends of the rod, which themselves are mounted in Al\textsubscript{2}O\textsubscript{3} centring plates providing mechanical support. These centring plates (coloured in light blue in Fig.~\ref{fig:schematicpaultrap}) are connected to OFHC copper brackets that are directly mounted to the roof of the inner heat shield as seen in Fig.~\ref{fig:schematicpaultrap}. Furthermore, the Paul trap is surrounded by a stainless steel cylinder attached to these brackets, providing additional stability for the mounting. In order to allow for optical access to the trapped ions as well as electrical access to the trap electrodes, the cylinder has openings of \SI{100}{\milli\meter} length at both sides and on the top of the trap as well as a small opening with a length of \SI{30}{\milli\meter} at its bottom. The electrical connection of the electrodes is provided through \SI{2}{\milli\meter} diameter copper wires directly soldered to them. These wires are pressed to the top of the inner heat shield in groups of seven with the use of clamps made from Al\textsubscript{2}O\textsubscript{3}, providing thermal contact to the heat shield while simultaneously keeping the wires electrically insulated from it. To provide electrical access to the trap electrodes, Kapton-insulated wires are soldered to the ends of the copper wires at the clamping point and connected to vacuum-compatible D-Sub-9 feedthroughs integrated into the wall of the inner heat shield. The connection to the electrical feedthroughs of the outer heat shield and the vacuum chamber is provided through vacuum compatible flat-ribbon cables.	A photograph of the trap assembly is shown in Fig.~\ref{fig:paultrapfoto}.

Similar to the other quadrupoles in the setup, the RF signal for the Paul trap is provided by a generator developed at Justus-Liebig-Universität Gießen. Due to significant capacitances in the cables and the RF-DC mixer, the  LC circuit has its resonance frequency at around \SI{2}{\mega\hertz} with an RF amplitude of up to \SI{900}{\volt pp} if the RF is applied only to two diagonally opposing rods, while the other pair of rods are grounded. In a balanced scheme with all four rods contributing to the overall capacitance, the resonance frequency is reduced to around \SI{1.5}{\mega\hertz}. 

For the creation of a confining potential along the ion axis, each of the seven trap electrode segments can be supplied with a DC voltage up to \SI{400}{\volt} by a PC controlled precision power supply (\textit{HV400}, Stahl-Electronics). The voltage stability provided by the HV400 is on the order of a few \num{e-6} relative to full scale. In case of the central trap segment, the voltage applied to each of the four electrodes of this segment can be controlled separately to allow for micromotion compensation of trapped ions. All voltages applied to the trap segments can be switched on and off via fast HV switches provided by Justus-Liebig-Universität Gießen. The switches are triggered by TTL signals from the output ports of a digital I/O card (\textit{NI-9401}, National Instruments) in the cRIO-9064 FPGA module and are controlled by the LabVIEW software running on the laboratory PC.

\begin{figure}[htb]
    \centering
    \includegraphics[width=\linewidth]{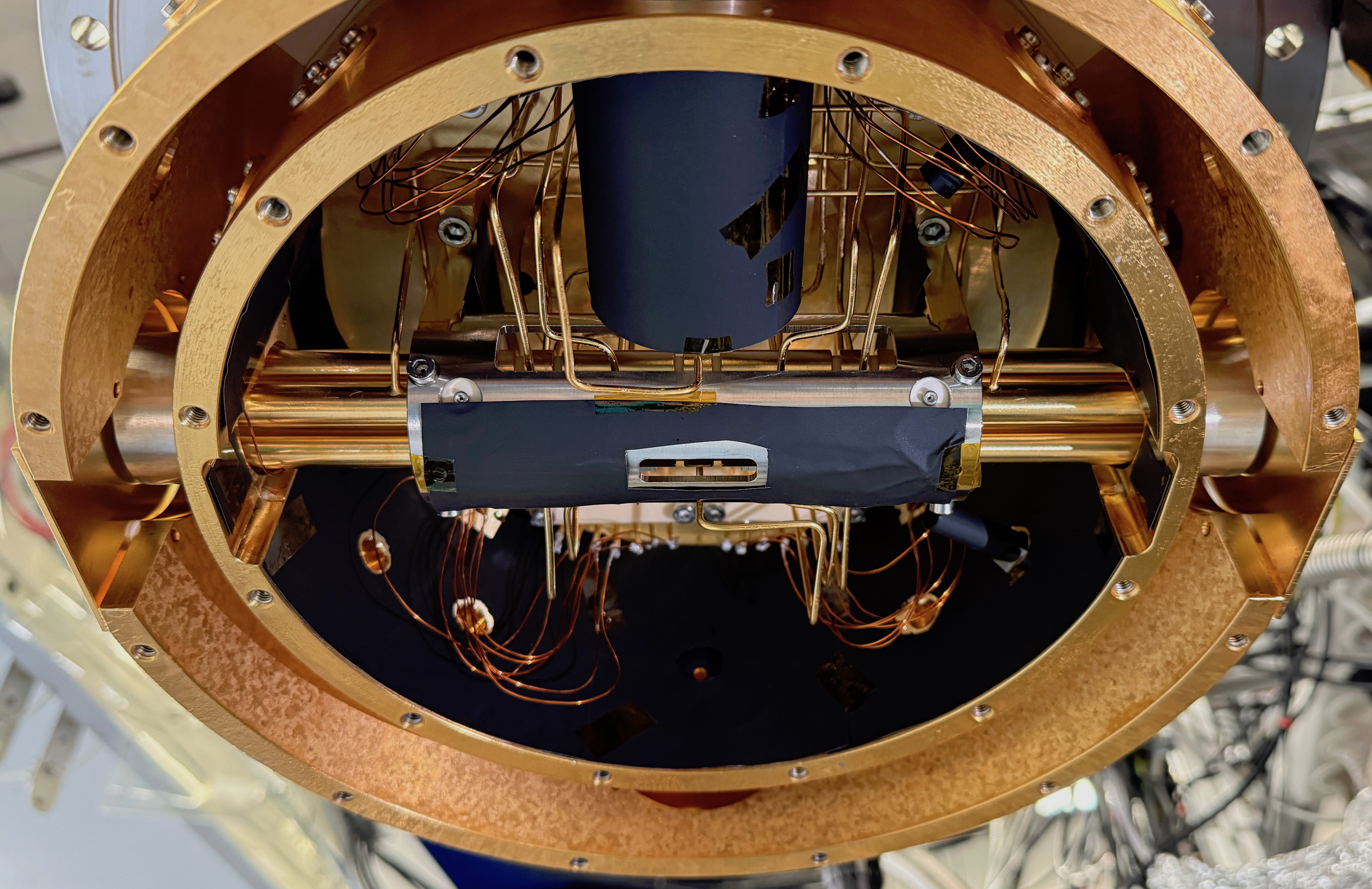}
    \includegraphics[width=\linewidth]{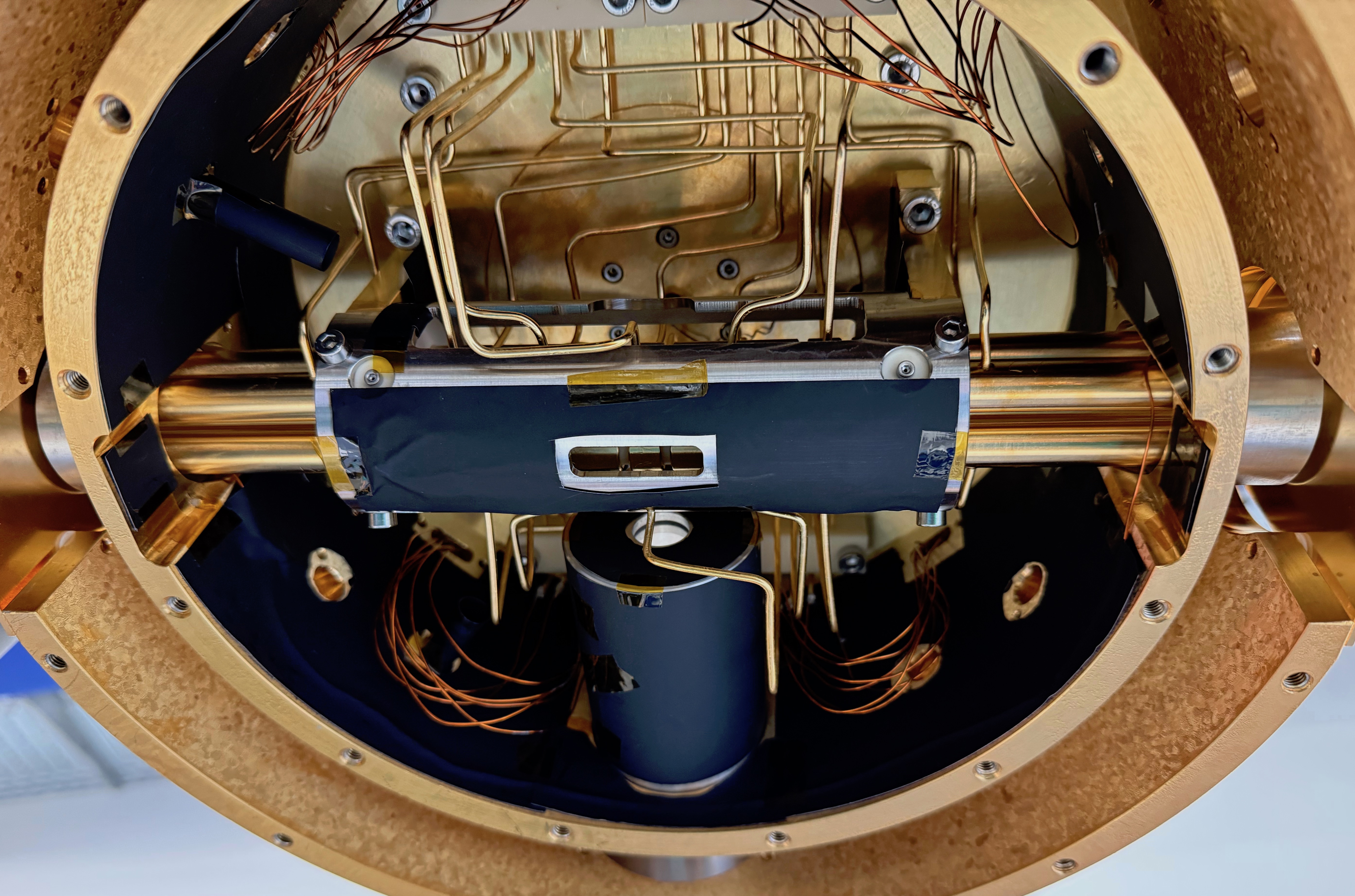}
    \caption{The inner surface of the trap chamber coated with black anodized aluminum foil for stray light absorption.}
    \label{fig:paultrapstraylight}
\end{figure}

At a later time, the inner walls of the \SI{4}{\kelvin} shield together with some of the other mechanical parts located inside of the shielding were covered with black anodized aluminum foil, which is shown in Fig.~\ref{fig:paultrapstraylight}. These measures were necessary to reduce the stray light from the laser beams that are guided through the electrode rods into the trap region (e.g. laser incoupling 1 in Fig.~\ref{fig:setupoverview}). For these beams, a reduction of stray light intensity by more than two orders of magnitude was observed.

\subsection{Ion detection}
\label{sec: Ion Detection}
For the detection of ions produced in the ion sources or ejected from the Paul trap and transmitted through the beam line, either an MCP or a CEM detector is mounted on axis in the detector chamber after QMS 1. The MCP consists of two plates in chevron configuration (GIDS GmbH) mounted in a commercial assembly (\textit{F9890-13}, Hamamatsu Photonics) with \SI{-1.8}{\kilo\volt} applied to the front plate. Signals from ionic impacts are read out via an anode after the back plate and further amplified in a modified fast-timing preamplifier (\textit{VT 120}, Ortec/Ametek).

Alternatively, a CEM (\textit{KBL15RS}, Dr. Sjuts Optomechanik) is used for the monitoring of transmitted ions. It is operated with \SI{-2.4}{\kilo\volt} bias voltage and read out either with the same Ortec preamplifier or with a low-noise amplifier (\textit{ZFL-1000LN+}, MiniCircuits).
Due to the compact dimensions of the CEM, it has been mounted on a linear mechanical feedthrough and can be moved several millimeters in the vertical direction to unblock the ion axis for laser access.
The amplified signals of both detector options are converted to TTL signals by a constant fraction discriminator and fed to the digital input of the NI-9401 card in the cRIO-9064 FPGA module for data processing on the laboratory PC.

\subsection{Vacuum system}
\label{sec:VacuumSystem}
The vacuum system of the experimental apparatus has to fulfill two main requirements:
\begin{itemize}
    \item The extraction of \textsuperscript{229(m)}Th\textsuperscript{3+} requires UHV conditions in the buffer-gas stopping cell and the adjacent Extraction RFQ to suppress contamination of the buffer gas during operation.
    \item While the buffer-gas stopping cell is operated at \SI{32}{\milli\bar}, the pressure in the Paul trap needs to be lower than \SI{e-8}{\milli\bar} to allow for formation of Coulomb crystals and even lower to achieve the long storage times needed for lifetime measurements. Therefore, differential pumping is crucial in order to prevent helium from reaching the trap.
\end{itemize}

\begin{figure*}[t]
	\centering
	\includegraphics[width = \textwidth]{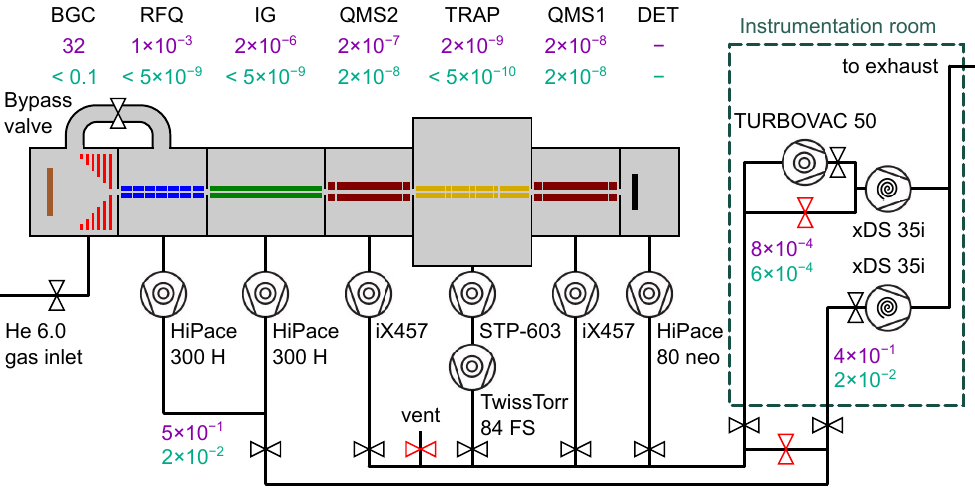}
	\caption{Pumping schematic. Red valves are normally closed. Pressures (in \si{\milli\bar}) during operation of the buffer-gas cell are shown in purple (inlet valve open and bypass valve closed). Background pressures (in \si{\milli\bar}) are shown in turquoise (inlet valve closed and bypass valve open).  BGC: buffer-gas cell. RFQ: radio-frequency quadrupole. IG: ion guide. QMS: quadrupole mass separator. DET: detector.  \label{fig:pumpoverview}}
\end{figure*}

An overview of the vacuum system is shown in Fig.~\ref{fig:pumpoverview}.
The first stage of differential pumping is provided by the chamber housing the Extraction RFQ, which is connected to the buffer-gas cell only through the de Laval nozzle with its throat diameter of \SI{0.4}{\milli\meter}. The chamber is equipped with a turbo-molecular pump (\textit{HiPace 300 H}, Pfeiffer Vacuum) with a pumping speed of \SI{255}{\litre\per\second} \cite{PfeifferHiPacePump} for helium, which is backed by a dry scroll pump (\textit{XDS35i}, Edwards Japan) with a pumping speed of \SI{35}{\cubic\meter\per\hour} \cite{Edwardsxds35i}. During operation, the buffer-gas stopping cell is kept at \SI{32}{\milli\bar}, resulting in a typical pressure of \SI{e-3}{\milli\bar} in the chamber of the Extraction RFQ.
In order to achieve the desired UHV conditions in the stopping cell and the chamber housing the Extraction RFQ, special care was taken to only use UHV-compatible materials for all components therein. Furthermore, the chambers are equipped to be baked out at a temperature of up to \SI{130}{\degreeCelsius} via a computer controlled heating system.
This system consists of twelve heating cartridges (\textit{Typ. HS-N 454}, HS Heizelemente) providing a heating power of \SI{80}{\watt} per cartridge. The cartridges have a length of \SI{40}{\milli\meter} and a diameter of \SI{4.5}{\milli\meter}. Eight of the cartridges are fitted into copper cylinders with an outer diameter of \SI{10}{\milli\meter} and lengths of \SI{40}{\milli\meter} and inserted into corresponding drill-holes in the corners of the cubic vacuum housing of the Extraction RFQ chamber. The remaining four cartridges are inserted into copper blocks with dimensions of $l \cdot w \cdot h = \SI{70}{\milli\meter} \cdot \SI{50}{\milli\meter} \cdot \SI{8}{\milli\meter}$ attached to the octagonal vacuum housing of the buffer-gas cell with thermally conducting glueing pads. Using the built-in thermocouples of the cartridges, their temperature is actively regulated through laboratory power supplies remotely controlled by an Arduino microcontroller. After the bake out process, base pressures below \SI{5e-9}{\milli\bar} are reached in the buffer-gas stopping cell and the Extraction RFQ chamber. In order to be able to effectively pump the buffer-gas stopping cell during the bake out process, it is connected to the Extraction RFQ chamber through a bypass equipped with a VAT angle valve as shown in Fig.~\ref{fig:stoppcelloverview}. The bypass as well as the CF100 tube connecting the turbo-molecular pump to the Extraction RFQ chamber are heated with silicone heating tape (\textit{Isopad IT-SiS10}, TC-E B.V.), providing temperatures up to \SI{200}{\degreeCelsius}. Their heating power is regulated through analog power regulators (\textit{Kemo FG002N}, Kemo Electronic).

The next two stages of differential pumping are provided by the Ion Guide chamber, which is connected to the Extraction RFQ chamber through an aperture opening of \SI{2}{\milli\meter} and to the subsequent QMS chamber by an aperture opening of \SI{3}{\milli\meter}, respectively. The Ion Guide chamber is pumped using a turbo-molecular pump (\textit{HiPace 300 H}, Pfeiffer Vacuum), which is backed by the same roughing pump as the turbo-molecular pump of the Extraction RFQ chamber. The resulting pressure in this chamber during buffer-gas operation is typically in the range of \SI{2e-6}{\milli\bar}.

The last differential stage before the trap chamber is the vacuum chamber of QMS 2 equipped with a turbo-molecular pump (\textit{iX457}, Edwards Japan) with a pumping speed of \SI{300}{\litre\per\second} \cite{EdwardsiX457} backed by a second roughing pump (\textit{XDS35i}, Edwards Japan). The vacuum chamber of QMS 2 is connected to the central chamber of the Paul trap through a \SI{5}{\milli\meter} aperture. During buffer-gas operation, the pressure in the QMS 2 chamber is \SI{2e-7}{\milli\bar}. The chamber housing QMS 1 is identical in construction to the one of QMS 2 also being connected to the trap chamber via an aperture with an opening of \SI{5}{\milli\meter}. It is pumped by a second turbo-molecular pump (\textit{iX457}, Edwards Japan) and backed by the same roughing pump. In order to compensate for the lower compression ratio for hydrogen compared to the HiPace 300 H, an additional turbo-molecular pump (\textit{Turbovac 50}, Leybold) can be added in series to the roughing pump. This results in a pre-vacuum in the \SI{e-4}{\milli\bar} range and to partial pressures of hydrogen below a few \SI{e-8}{\milli\bar} in the QMS chambers.

The central chamber housing the cryogenic linear Paul trap is pumped by a turbo-molecular pump (\textit{STP 603}, Edwards Japan), providing a pumping speed of \SI{650}{\litre\per\second} \cite{EdwardsSTP603}. The pump is operated in series with a second turbo-molecular pump (\textit{TwissTorr 84 FS}, Agilent Technologies) with a pumping speed of \SI{49}{\litre\per\second} \cite{AgilentPumpe} to achieve a lower partial pressure of hydrogen in the chamber. The Edwards turbo-molecular pump is directly attached to the bottom of the vacuum chamber. This is enabled by a circular hole of \SI{430}{\milli\metre} diameter in the optical table onto which the trap chamber is mounted. Both turbo-molecular pumps are backed by the same roughing pump (\textit{XDS35i}, Edwards Japan) that also serves the two turbo-molecular pumps that evacuate the quadrupole mass separators. Without operating the cold head, a pressure in the low \SI{e-9}{\milli\bar} region is achieved. When the cold head is operated, the pressure outside the \SI{40}{\kelvin} shield decreases below the measurable range of \SI{5e-10}{\milli\bar} of the pressure gauge (\textit{PBR 260}, Pfeiffer Vacuum) \cite{PfeifferPBR260} attached to the chamber. However, during operation of the buffer-gas cell, the pressure rises to about \SI{2e-9}{\milli\bar}.

The detector vacuum chamber is connected to the previous QMS 1 chamber via an aperture with \SI{8}{\milli\meter} diameter and is separately evacuated by a turbo-molecular pump (\textit{HiPace 80 Neo}, Pfeiffer Vacuum) with a pumping speed of \SI{58}{\litre\per\second} for helium \cite{PfeifferHiPace80neo} and as well backed by the same roughing pump of the quadrupole mass separator chambers and the trap chamber. For summary, the background pressures (turquoise) and the pressures during buffer-gas operation (purple) are also tabulated in Fig.~\ref{fig:pumpoverview}.

\section{Thorium extraction}
\label{sec: Thorium Extraction}

\begin{figure}[htb]
    \centering
    \includegraphics[width = \linewidth]{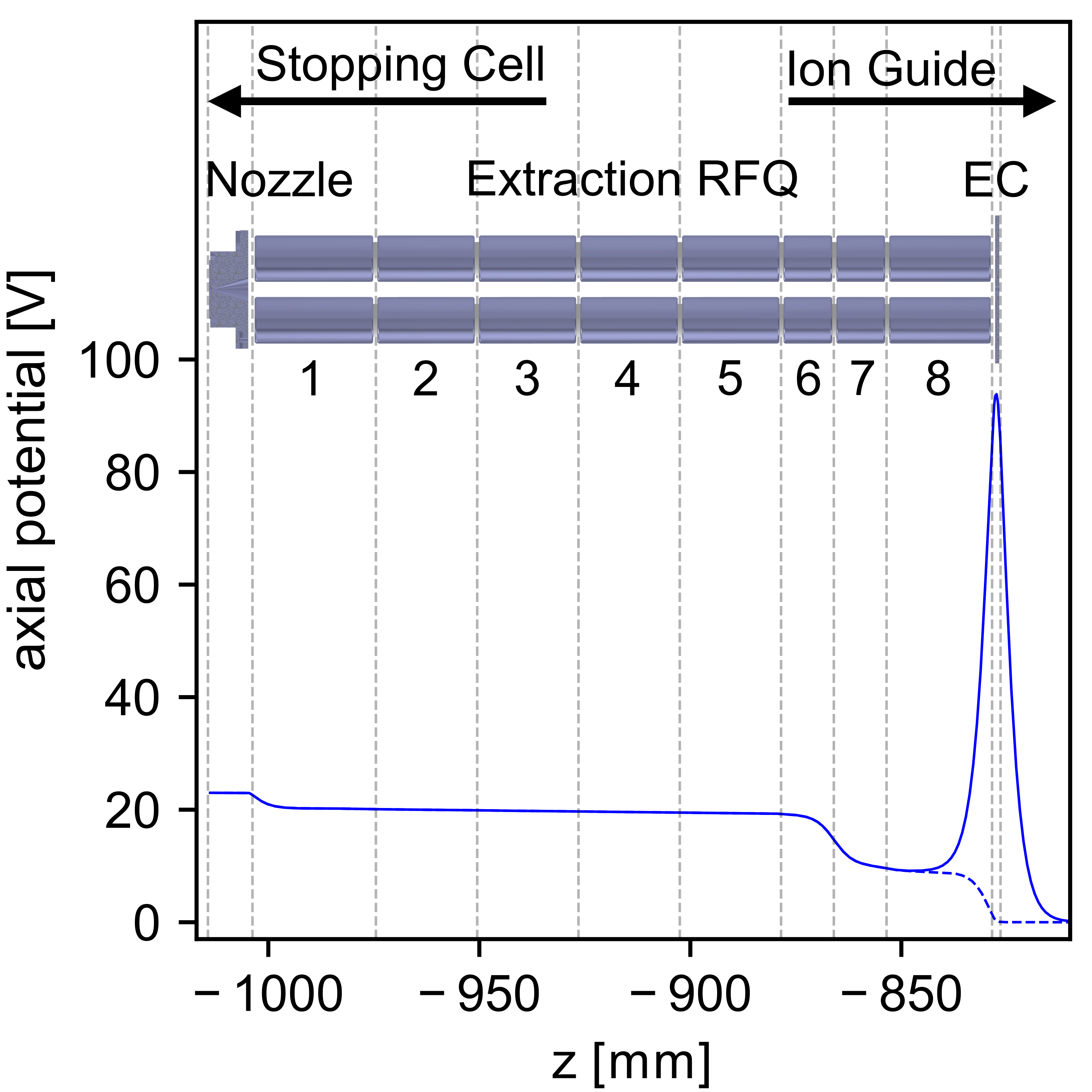}
	\caption{Potential along the ion axis from the supersonic nozzle until the switchable endcap electrode of the segmented Extraction RFQ with the actual experimental parameters applied to the different electrodes. The state of ion collection with the ramped up endcap voltage is shown as a solid line, and the state of bunch release as a dashed line. Both potential shapes were calculated with the SIMION software (Version 8.1.1.32) \cite{dahl2000simion}. 
    \label{fig: RFQ Axial Potential}}
\end{figure}

\begin{figure}[htb]
    \centering
    \includegraphics[width=\linewidth]{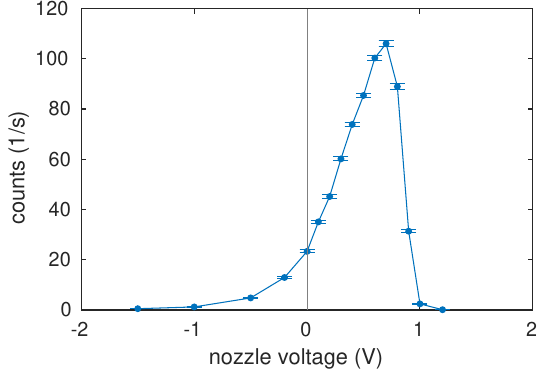}
    \caption{Number of extracted ions as a function of nozzle voltage. The nozzle voltage is measured relative to the voltage at the exit of the funnel (vertical line).}
    \label{fig:nozzlevoltage}
\end{figure}

For extraction of thorium ions from the buffer-gas stopping cell, we start with the adjustment of the pressure of the ultra-pure helium to \SI{32}{\milli\bar}. The \textsuperscript{233}U source is set to a DC offset of \SI{42.0}{\volt}, the last segment of the funnel to \SI{22.5}{\volt}, and the first segment of the Extraction RFQ to \SI{20.2}{\volt}. For continuous extraction, the voltage on each subsequent segment of the Extraction RFQ is reduced by \SI{0.2}{\volt}, see dashed line in Fig.~\ref{fig: RFQ Axial Potential}. These values are comparable to the ones used in previous experiments \cite{Wense15}. Stopped ions are guided via the RF-DC-funnel through the de Laval nozzle to the Extraction RFQ. Typically, the nozzle is set to a DC voltage of about \SI{0.7}{\volt} above the last funnel electrode (which implies \SI{23.2}{\volt} absolute voltage) to provide optimum ion transmission. 
We observe that the transmission is very sensitive to the nozzle voltage, and a nozzle voltage scan is performed after each time the other parameters in the buffer-gas cell or Extraction RFQ are changed, see Fig.~\ref{fig:nozzlevoltage} for a typical scan.

\begin{figure*}[t]
	\centering
	\includegraphics[width = 1\textwidth]{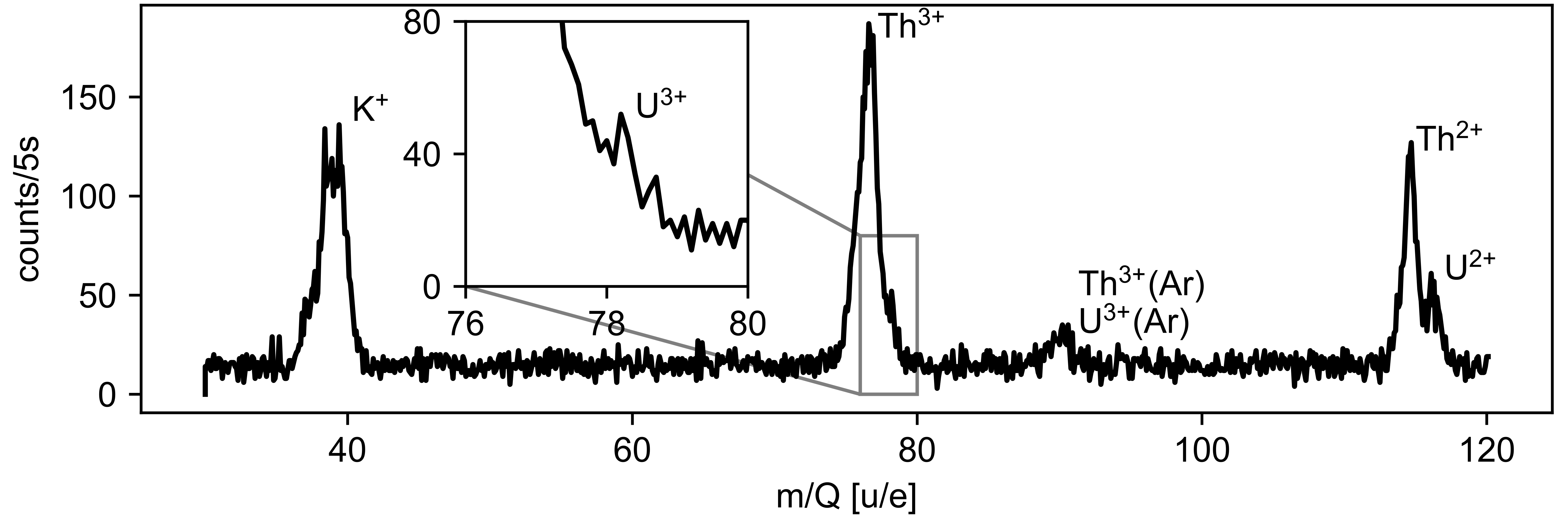}
	\caption{Mass scan of continuously extracted ions from the buffer-gas stopping cell performed with QMS 1. The filter width is $\Delta(m/Q) = \SI{2}{u/e}$ and the integration time is \SI{5}{\second} per scan step of \SI{0.1}{u/e}. \label{fig: Th MassScan}}
\end{figure*}

A mass scan of the ions extracted from the buffer-gas stopping cell performed with QMS 1 is shown in Fig.~\ref{fig: Th MassScan}. Here, the Extraction RFQ, the Ion Guide, the QMS 2, as well as the Paul trap, were set to transmit all masses above $m = \SI{30}{u/e}$. Compared to a test measurement, in which the signal has been recorded as close as possible to the output of the buffer-gas cell, here, the detector is positioned roughly \SI{1.4}{\meter} after the output of the buffer-gas cell. Consequently, the long drift length influences the mass resolution. For characterization of the content of the bunches extracted from the buffer-gas cell, the use of either QMS 1 or QMS 2 is possible. In our case, QMS 1 turned out to have a better mass resolution than QMS 2.

Apart from the main peaks originating from \textsuperscript{229(m)}Th\textsuperscript{3+}, \textsuperscript{233}U\textsuperscript{3+}, \textsuperscript{229(m)}Th\textsuperscript{2+} and \textsuperscript{233}U\textsuperscript{2+} with their respective mass to charge ratios of \SI{76.3}{u/e}, \SI{77.7}{u/e}, \SI{114.5}{u/e} and \SI{116.5}{u/e}, there is also a major contribution at \SI{39}{u/e}. Most probably, it results from a contamination of the funnel electrodes with potassium during the calibration of the QMS with a heated ion source. 
In comparison with the mass spectra measured with the previously used large LMU buffer-gas stopping cell \cite{Wense15, WenseThesis, Nature16Wense} the count rates at the same integration time are reduced by a factor of 20 to 200. This can be mainly explained by the lower activity of the \textsuperscript{233}U source of only \SI{10}{\kilo\becquerel} compared to the \SI{290}{\kilo\becquerel} in the old setup in combination with a similar dark count rate of the detector. Unfortunately, the separability of neighbouring peaks (especially in the case of \textsuperscript{229(m)}Th\textsuperscript{3+} and \textsuperscript{233}U\textsuperscript{3+}) is somewhat worse than in previous experiments. A possible reason is that the long beam line leads to additional instability compared to previous experiments.

For bunched extraction from the Extraction RFQ the optimal DC voltages turned out to be \SI{20.2}{\volt} to \SI{19.2}{\volt} in steps of \SI{0.2}{\volt} for the segments 1 to 6 as well as \SI{10.0}{\volt} and \SI{9.0}{\volt} for the segments 7 and 8, respectively, see Fig.~\ref{fig: RFQ Axial Potential}. The RFQ endcap aperture is switched from \SI{100}{\volt} applied during accumulation to \SI{0}{\volt} to release the bunches. During the accumulation in segment 8, thorium ions are phase-space cooled to room temperature in the \SI{e-3}{\milli\bar} helium buffer gas. Usually, thorium ions are accumulated for \SI{10}{\second} to create a bunch containing several hundred ions.

For the extraction of a pure bunch of \textsuperscript{229(m)}Th\textsuperscript{3+} ions, the Ion Guide is used as a first coarse mass filter for ions of mass to charge ratio of \SI{75}{u/e}, which corresponds to the rising edge of the \textsuperscript{229(m)}Th\textsuperscript{3+} peak in the mass scan. Operational parameters for this filtering are an RF amplitude of \SI{172}{\volt pp} and a DC voltage of \SI{\pm 11.7}{\volt}, resulting in the Mathieu parameters of $a = 0.104$ and $q = 0.766$.
More precise filtering is then undertaken with the adjoining QMS 2 for the same mass to charge ratio and a smaller filter width of choice. In such a configuration, the RF amplitude is \SI{605}{\volt pp} and the applied filtering DC voltage \SI{\pm 49.3}{\volt} resulting in Mathieu parameters of $a = 0.112$ and $q =0.688$.
The pure \textsuperscript{229(m)}Th\textsuperscript{3+} ion bunch can then be used for Paul trap experiments.

\section{Sympathetic laser cooling and fluorescence detection}
\label{sec: Proof of Principle Measurements}

The starting condition for experiments on spectroscopy of the \textsuperscript{229(m)}Th\textsuperscript{3+} electronic hyperfine structure or the direct VUV excitation of the thorium nucleus is the stable trapping of a laser-cooled \textsuperscript{88}Sr\textsuperscript{+} - \textsuperscript{229(m)}Th\textsuperscript{3+} mixed-species Coulomb crystal.

For sympathetic cooling \textsuperscript{88}Sr\textsuperscript{+} was chosen because it has similar charge-to-mass ratio as \textsuperscript{229(m)}Th\textsuperscript{3+}, it has no hyperfine structure, and it can be comfortably and effectively laser cooled using two diode lasers at wavelengths of \SI{422}{\nano\meter} and \SI{1092}{\nano\meter} \cite{removille2009trapping, jung2017all} that drive transitions of $\approx \SI{20}{\mega\hertz}$ linewidth \cite{gallagher1967oscillator}.
The wavelength of these lasers provides sufficient separation from the \SI{690}{\nano\meter} resonance transition 5f F$_{5/2}$ $\rightarrow$ 6d D$_{5/2}$ of \textsuperscript{229(m)}Th\textsuperscript{3+} used for fluorescence detection of the isomer \cite{ThoriumnuclearClockPI, zitzer2024sympathetic}. 

\begin{figure}[htb]
    \centering
    \includegraphics[width=\linewidth]{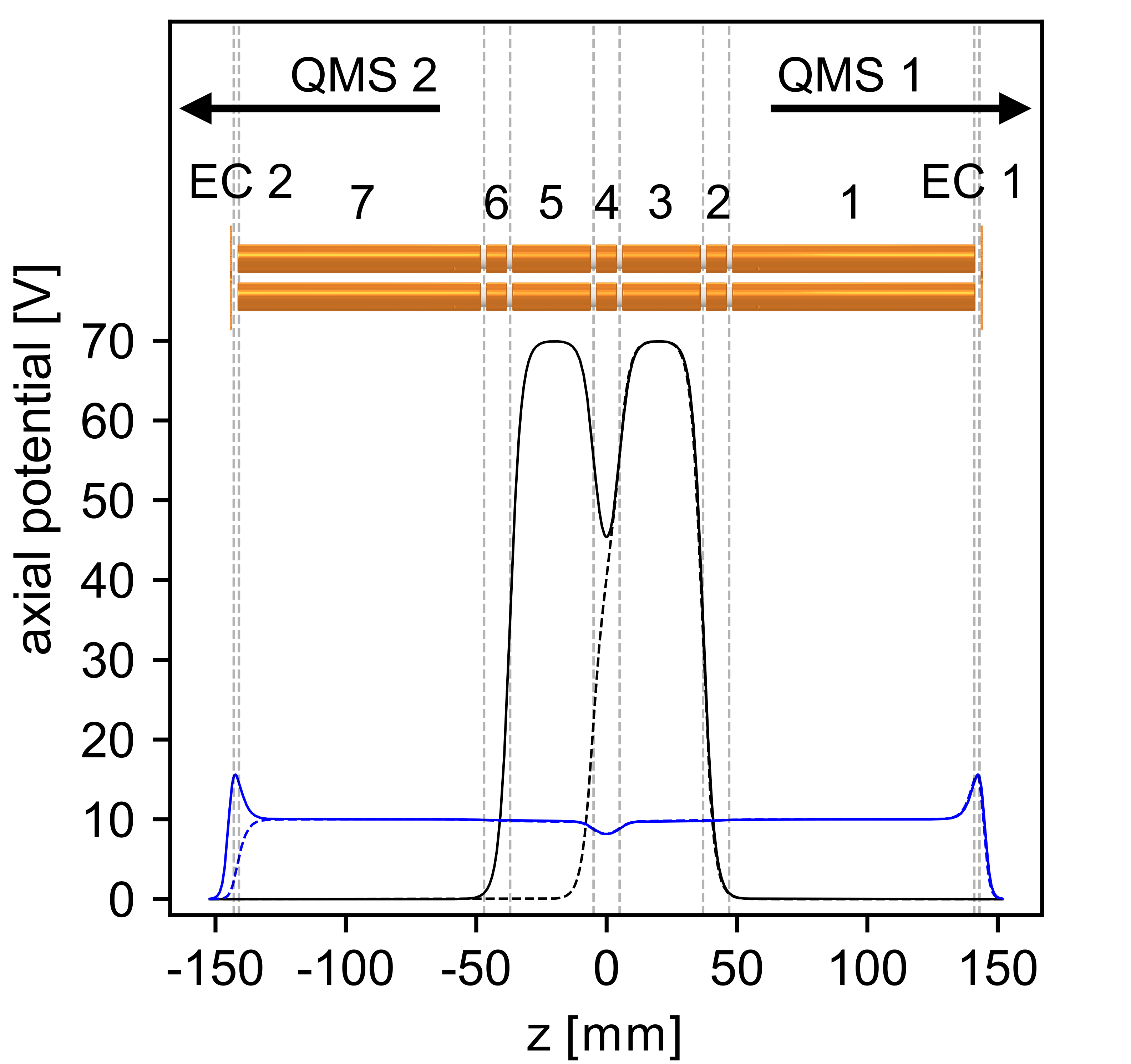}
    \caption{Trapping potentials along the ion axis for the catching (dashed lines) and confinement (solid lines) of ablated Sr ions (black) and for the capture of \textsuperscript{229}Th\textsuperscript{3+} ions extracted from the buffer-gas stopping cell (blue). The potential shapes are dictated by the voltages applied to the seven trap electrode segments as well as the endcap apertures (EC) and were calculated with the SIMION simulation software (Version 8.1.1.32) \cite{dahl2000simion}.}
    \label{fig:Trap potentials}
\end{figure}

To create a Coulomb crystal within the LMU setup, the following routine is executed: The Paul trap is operated in an unbalanced mode with the RF voltage applied only to one electrode pair and the other electrode pair grounded. In this configuration, the resonance frequency is pushed closer to \SI{2}{\mega\hertz} and higher RF amplitudes of approximately \SI{900}{\volt pp} can be achieved. By applying DC voltages of \SI{40}{\volt} to the central trap segment (segment 4) and \SI{70}{\volt} to segment 3, respectively, a blocking axial potential is created, see black dashed line in Fig.~\ref{fig:Trap potentials}. After a delay of \SI{100}{\micro\second} with respect to a pulse of the ablation laser, \SI{70}{\volt} are applied to segment 5 to form the \textsuperscript{88}Sr\textsuperscript{+} confinement potential, see black line in Fig.~\ref{fig:Trap potentials}. Only a small fraction of the velocity distribution of the ablated ions is trapped in this way. However, several hundred mass-filtered \textsuperscript{88}Sr\textsuperscript{+} ions can be confined, which is sufficient for the desired number of co-trapped thorium ions.

Doppler cooling of the \textsuperscript{88}Sr\textsuperscript{+} ions is done with in total of 4 laser beams collinearly aligned through one of the viewports at \SI{30}{\degree} with respect to the ion axis. The main cooling transition is driven at \SI{422}{\nano\meter} with a power of \SI{300}{\micro\watt} and a detuning of \SI{-10}{\mega\hertz}. Two additional \SI{422}{\nano\meter} laser beams which are detuned by \SI{-200}{\mega\hertz} and \SI{-400}{\mega\hertz} with respect to the cooling transition act as "catcher beams" and improve the initial \textsuperscript{88}Sr\textsuperscript{+} cooling process as well as the conservation of the crystal during the following \textsuperscript{229(m)}Th\textsuperscript{3+} ion capture. The fourth beam for repumping at \SI{1092}{\nano\meter} with a power of \SI{700}{\micro\watt} prevents the pumping into a dark state. For more details on the laser setup, see \cite{ScharlSetup2023}.
A \textsuperscript{88}Sr\textsuperscript{+} Coulomb crystal typically forms after a few seconds and can be observed on the EM-CCD camera via the fluorescence readout at \SI{422}{\nano\meter}.

After a successful \textsuperscript{88}Sr\textsuperscript{+} cooling and crystallization, the axial potential is lowered and reshaped into a double-well for thorium ion loading (depicted in blue in Fig.~\ref{fig:Trap potentials}).
In parallel, the buffer-gas stopping cell is operated in a \textit{stand by} mode with the endcap aperture of the Extraction RFQ ramped up to \SI{100}{\volt} for constant collection of thorium ions. Following the preparation of the double-well axial potential, a single thorium ion bunch is then released from the  Extraction RFQ, which contains roughly 100 pure \textsuperscript{229(m)}Th\textsuperscript{3+} ions after mass purification in the QMS 2 (or Ion Guide). Before the release of the thorium ion bunch, the trap endcap voltage applied to EC 2 is switched off and is ramped up again \SI{270}{\micro\second} after the bunch release. The \textsuperscript{229(m)}Th\textsuperscript{3+} ions are initially confined in the bathtub-shaped outer potential well spanning the whole trap length. As they interact with the continuously cooled \textsuperscript{88}Sr\textsuperscript{+} ions in the trap center, they are significantly slowed down and are confined in the central potential well. There, a compound Coulomb crystal containing both \textsuperscript{88}Sr\textsuperscript{+} and \textsuperscript{229(m)}Th\textsuperscript{3+} ions is created and can be further reshaped by changing the axial potential.

\begin{figure}[htb]
    \centering
    \includegraphics[width=\linewidth]{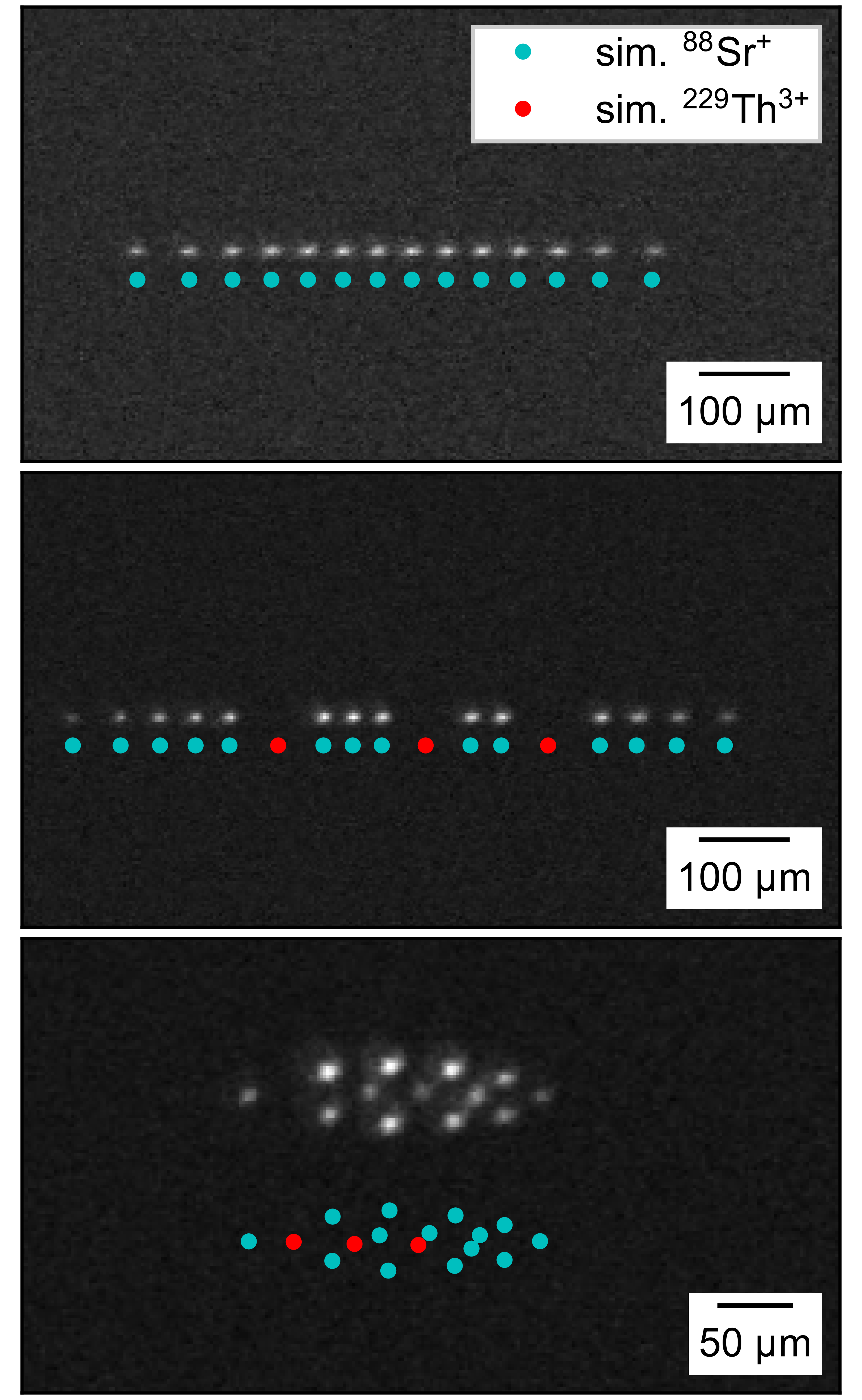}
    \caption{EM-CCD images of the same 14-ion \textsuperscript{88}Sr\textsuperscript{+} Coulomb crystal: before (upper image) and after (central image) the implantation of three \textsuperscript{229(m)}Th\textsuperscript{3+} ions within an axial potential depth of \SI{0.1}{\volt}. The lower image shows the 3D crystal formed at an axial potential depth of \SI{3.0}{\volt}. For visualization of the thorium ion positions, SIMION (Version 8.1 for the upper and center images and 8.0 for lower image) Coulomb crystal simulations are added \cite{dahl2000simion}. The integration time is \SI{1.0}{\second} for all images.
    \label{fig:Sr-Th Crystal}}
\end{figure}

In Fig.~\ref{fig:Sr-Th Crystal}, an example of an initial 14-ion \textsuperscript{88}Sr\textsuperscript{+} Coulomb crystal is given together with two recordings after the implantation of three \textsuperscript{229(m)}Th\textsuperscript{3+} ions. The linear ion chains in the upper two pictures were recorded with a potential depth of \SI{0.1}{\volt}. To realize the three-dimensional crystal, a potential depth of \SI{3.0}{\volt} was applied.
Only the \SI{422}{\nano\meter} fluorescence of \textsuperscript{88}Sr\textsuperscript{+} ions was detected. Consequently, the \textsuperscript{229(m)}Th\textsuperscript{3+} ions are only visible as dark spots. To visualize the positions of the \textsuperscript{229(m)}Th\textsuperscript{3+} ions and to get a better understanding of the size and shape of the Coulomb crystal, SIMION \cite{dahl2000simion} simulations were performed using the same trapping voltage settings as in the experiment. The scale bars in Fig.~\ref{fig:Sr-Th Crystal} are based on the fits of the image magnification to the simulations. Suitable fits for the image magnification turned out to be 5.35 and 5.335 in case of the two linear chains and 5.6 for the 3D crystal. Estimating the magnification of the used lens system can be done by looking at the object width $d_o$ and image width $d_i$ resulting in the optimum resolved image on the camera. The used lens system for these experiments is slightly altered compared to the imaging setup described in \cite{ScharlSetup2023}. It now only consists of a $f = \SI{50}{\milli\meter}$ aspheric lens (\textit{25mm Dia 0.25 NA UV-VIS Coated, UV Fused Silica Aspheric Lens, 33-958}, Edmund Optics) for photon collection in vacuum and a $f = \SI{300}{\milli\meter}$ plano-convex lens (\textit{LA1484}, Thorlabs) in air for focusing of the collimated fluorescence onto the sensor of the EM-CCD camera. Approximating the image magnification with $M = d_i/d_o = \SI{243.7(5)}{\milli\meter}/\SI{45.9(5)}{\milli\meter} = 5.31(6)$ is in good agreement with the simulation of the linear ion chains but deviates by \SI{5}{\percent} for the 3D crystal. Given the uncertainty of the calibration of the voltage sources and the approximations used in the simulation, this is still a reasonable agreement. Consequently, we find that the choice of the SIMION simulation software is appropriate for the modelling of linear chains and three-dimensional Coulomb crystals, and can provide accurate positions of the \textsuperscript{229(m)}Th\textsuperscript{3+} ions within the mixed-species Coulomb crystals.

The entire trapping routine is automated to facilitate upcoming spectroscopy measurements and to ensure reproducibility of the \textsuperscript{88}Sr\textsuperscript{+} - \textsuperscript{229(m)}Th\textsuperscript{3+} mixed-species Coulomb crystal preparation process. As discussed in \cite{ScharlSetup2023}, the envisaged scheme for the thorium isomeric lifetime measurement using single \textsuperscript{229m}Th\textsuperscript{3+} ions has to be repeated many times in order to achieve low statistical uncertainty.

\section{Vacuum characterization in the trap region}
\label{sec: Vacuum Characterization in the Trap Region}
To check the suitability of our cryogenic Paul trap for lifetime measurements of the \textsuperscript{229m}Th\textsuperscript{3+} isomer, we characterized the ion storage time using \textsuperscript{88}Sr\textsuperscript{+} ions. We observed the fluorescence of 9 trapped \textsuperscript{88}Sr\textsuperscript{+} ions over 8 days. During this period, 3 fluorescence decays could be observed resulting from chemical reactions of the \textsuperscript{88}Sr\textsuperscript{+} with residual particles in the trap. However, the ions were not lost, but instead formed \textsuperscript{88}SrH\textsuperscript{+} molecular ions. Consequently, lower limit for the storage time then lies around $\tau = \SI{192}{\hour}$, which well meets the requirements.

Since hot cathode ionization pressure gauges are not capable to measure pressures below \SI{e-12}{\milli\bar}, the only option is to use the fluorescence lifetime of a trapped ion as a measure for the vacuum pressure inside the cryogenic trap volume. As already shown by other research groups working with cryogenic ion traps, an estimate for the vacuum pressure can be given using the Langevin collision model \cite{schmoeger2015coulomb, schmoeger2017kalte, pagano2019cryogenic}. Under the assumption of H\textsubscript{2} as the only remaining gas-phase collision partner at temperatures of \SI{8}{\kelvin} in the Paul trap and following the descriptions in 
\cite[p.168 --180]{schmoeger2017kalte}, the particle density of hydrogen is given in dependence of the Langevin rate coefficient $k_L$. This coefficient, in turn, is only defined by the charge state of the ion, the polarizability of the hydrogen molecules, and the reduced mass of the two collision partners, thus
\begin{equation}
	n_{H_2} = \frac{1}{\tau k_L} \approx \SI{9.7e+8}{1/m^3}.
\end{equation}
Further assuming an ideal gas model, the partial pressure of the hydrogen molecules $p_{H_2}$ in the trap volume can be estimated to
\begin{equation}
	p_{H_2}(\SI{8}{\kelvin}) = n_{H_2}k_BT \approx \SI{1.07e-15}{\milli\bar}. 
\end{equation}
The derived value is consistent with vacuum estimates for other cryogenic ion traps \cite{schwarz2012cryogenic,repp2012pentrap, sellner2017improved,schmoeger2017kalte, micke2019closed, pagano2019cryogenic,leopold2019cryogenic, Godun2014}.

\section{Conclusion}
We have presented a comprehensive and detailed account of the cryogenic Paul trap experiment at LMU Munich for trapping and sympathetic cooling of \textsuperscript{229(m)}Th\textsuperscript{3+} ions by laser-cooled \textsuperscript{88}Sr\textsuperscript{+} ions. Technical specifications and working parameters are provided as well as characterizations of the ion sources, the mass separators, and the Paul trap. With the successful and controlled loading of \textsuperscript{229(m)}Th\textsuperscript{3+} ions embedded in a \textsuperscript{88}Sr\textsuperscript{+} mixed-species Coulomb crystal, this setup will be a workhorse for upcoming experiments. Measurements of the \textsuperscript{229(m)}Th\textsuperscript{3+} hyperfine structure as necessary steps towards the determination of the \textsuperscript{229m}Th\textsuperscript{3+} isomeric lifetime in vacuum are to be undertaken in the near future. In the long run, this setup offers the prospect for direct VUV excitation of the nuclear excited state in \textsuperscript{229}Th\textsuperscript{3+} ions. It could therefore serve as platform for a prototype of a trap-based thorium nuclear clock.

\section*{Acknowledgements}	
The authors want to thank their colleagues Ekkehard Peik, Maksim Okhapkin, Johannes Tiedau, and Gregor Zitzer at PTB Braunschweig for fruitful discussions. Furthermore, we want to thank Peter Micke, Lisa Schmöger, and Maria Schwarz as former members of the CryPTEx team at MPIK in Heidelberg for sharing their expertise on cryogenic ion traps.

This work is part of the ‘ThoriumNuclearClock’ project that has received funding from the European Research Council (ERC) under the European Union’s Horizon 2020 Research and Innovation Program (Grant Agreement No. 856415). It was supported by DFG (Th956/3-1, Th956/3-2), by BACATEC (Grant 7 [2019-2]), by the European Union's Horizon 2020 Research and Innovation Program under grant agreement 664732 “nuClock” and by the Ludwig-Maximilians-Universität (LMU) Department of Medical Physics via the Maier-Leibnitz Laboratory.

\section*{Data Availability Statement}
The data presented in this manuscript is available on the following zenodo repository: \url{https://doi.org/10.5281/zenodo.16691405}.

\bibliographystyle{naturemag}
\bibliography{bibliography_paper}

\begin{thebibliography}{10}
\expandafter\ifx\csname url\endcsname\relax
  \def\url#1{\texttt{#1}}\fi
\expandafter\ifx\csname urlprefix\endcsname\relax\def\urlprefix{URL }\fi
\providecommand{\bibinfo}[2]{#2}
\providecommand{\eprint}[2][]{\url{#2}}

\bibitem{NNDCnudat}
\bibinfo{author}{{Brookhaven National Laboratory}}.
\newblock \bibinfo{title}{{National Nuclear Data Center: Chart of Nuclides}}.
\newblock \bibinfo{howpublished}{Website}.
\newblock \urlprefix\url{https://www.nndc.bnl.gov/nudat3/}.
\newblock \bibinfo{note}{(accessed on 04.12.2024)}.

\bibitem{Kroger76}
\bibinfo{author}{Kroger, L.~A.} \& \bibinfo{author}{Reich, C.~W.}
\newblock \bibinfo{title}{{Features of the low-energy level scheme of ${}^{229}$Th as observed in the $\alpha$-decay of $^{233}$U}}.
\newblock \emph{\bibinfo{journal}{Nucl. Phys. A}} \textbf{\bibinfo{volume}{259}}, \bibinfo{pages}{29--60} (\bibinfo{year}{1976}).

\bibitem{Nature16Wense}
\bibinfo{author}{von~der Wense, L.} \emph{et~al.}
\newblock \bibinfo{title}{{Direct detection of the ${}^{229}$Th nuclear clock transition}}.
\newblock \emph{\bibinfo{journal}{Nature}} \textbf{\bibinfo{volume}{533}}, \bibinfo{pages}{47--51} (\bibinfo{year}{2016}).

\bibitem{Nature23Kraemer}
\bibinfo{author}{Kraemer, S.} \emph{et~al.}
\newblock \bibinfo{title}{{Observation of the radiative decay of the ${}^{229}$Th nuclear clock isomer}}.
\newblock \emph{\bibinfo{journal}{Nature}} \textbf{\bibinfo{volume}{617}}, \bibinfo{pages}{706--710} (\bibinfo{year}{2023}).

\bibitem{PhysRevTiedauSchaden24}
\bibinfo{author}{Tiedau, J.} \emph{et~al.}
\newblock \bibinfo{title}{{Laser Excitation of the Th-229 Nucleus}}.
\newblock \emph{\bibinfo{journal}{Phys. Rev. Lett.}} \textbf{\bibinfo{volume}{132}}, \bibinfo{pages}{182501} (\bibinfo{year}{2024}).

\bibitem{elwellPhysRevLett24}
\bibinfo{author}{Elwell, R.} \emph{et~al.}
\newblock \bibinfo{title}{Laser excitation of the $^{229}\mathrm{Th}$ nuclear isomeric transition in a solid-state host}.
\newblock \emph{\bibinfo{journal}{Phys. Rev. Lett.}} \textbf{\bibinfo{volume}{133}}, \bibinfo{pages}{013201} (\bibinfo{year}{2024}).

\bibitem{zhang2024_nature}
\bibinfo{author}{Zhang, C.} \emph{et~al.}
\newblock \bibinfo{title}{{Frequency ratio of the ${}^{229m}$Th nuclear isomeric transition and the $^{87}$Sr atomic clock}}.
\newblock \emph{\bibinfo{journal}{Nature}} \textbf{\bibinfo{volume}{633}}, \bibinfo{pages}{63--70} (\bibinfo{year}{2024}).

\bibitem{Zhang2024ThF}
\bibinfo{author}{Zhang, C.} \emph{et~al.}
\newblock \bibinfo{title}{{$^{229}\text{ThF}_4$ thin films for solid-state nuclear clocks}}.
\newblock \emph{\bibinfo{journal}{Nature}} \textbf{\bibinfo{volume}{636}}, \bibinfo{pages}{603--608} (\bibinfo{year}{2024}).

\bibitem{PhyRevC15Tkalya}
\bibinfo{author}{Tkalya, E.~V.}, \bibinfo{author}{Schneider, C.}, \bibinfo{author}{Jeet, J.} \& \bibinfo{author}{Hudson, E.~R.}
\newblock \bibinfo{title}{{Radiative lifetime and energy of the low-energy isomeric level in $^{229}$Th}}.
\newblock \emph{\bibinfo{journal}{Phys. Rev. C}} \textbf{\bibinfo{volume}{92}}, \bibinfo{pages}{054324} (\bibinfo{year}{2015}).

\bibitem{NatureYamaguchi24}
\bibinfo{author}{Yamaguchi, A.} \emph{et~al.}
\newblock \bibinfo{title}{{Laser spectroscopy of triply charged ${}^{229}$Th isomer for a nuclear clock}}.
\newblock \emph{\bibinfo{journal}{Nature}} \textbf{\bibinfo{volume}{629}}, \bibinfo{pages}{62--66} (\bibinfo{year}{2024}).

\bibitem{PhysScr96Tkalya}
\bibinfo{author}{Tkalya, E.~V.}, \bibinfo{author}{Varlamov, V.~O.}, \bibinfo{author}{Lomonosov, V.~V.} \& \bibinfo{author}{Nikulin, S.~A.}
\newblock \bibinfo{title}{{Processes of the Nuclear Isomer ${}^{229m}$Th ($3/2^+,3.5\pm1.0$eV) Resonant Excitation by Optical Photons}}.
\newblock \emph{\bibinfo{journal}{Phys. Scr.}} \textbf{\bibinfo{volume}{53}}, \bibinfo{pages}{296--299} (\bibinfo{year}{1996}).

\bibitem{PhysLett03Peik}
\bibinfo{author}{Peik, E.} \& \bibinfo{author}{Tamm, C.}
\newblock \bibinfo{title}{{Nuclear laser spectroscopy of the \SI{3.5}{\electronvolt} transition in ${}^{229}$Th}}.
\newblock \emph{\bibinfo{journal}{Europhys. Lett.}} \textbf{\bibinfo{volume}{61}}, \bibinfo{pages}{181--186} (\bibinfo{year}{2003}).

\bibitem{PhysRev12Campbell}
\bibinfo{author}{Campbell, C.~J.} \emph{et~al.}
\newblock \bibinfo{title}{{Single-Ion nuclear clock for metrology at the 19th decimal place}}.
\newblock \emph{\bibinfo{journal}{Phys. Rev. Lett.}} \textbf{\bibinfo{volume}{108}}, \bibinfo{pages}{120802} (\bibinfo{year}{2012}).

\bibitem{Brewer19}
\bibinfo{author}{Brewer, S.~M.} \emph{et~al.}
\newblock \bibinfo{title}{{$^{27}$Al$^+$ Quantum-Logic Clock with a Systematic Uncertainty below $10^{-18}$}}.
\newblock \emph{\bibinfo{journal}{Phys. Rev. Lett.}} \textbf{\bibinfo{volume}{123}}, \bibinfo{pages}{033201} (\bibinfo{year}{2019}).

\bibitem{Aeppli2024}
\bibinfo{author}{Aeppli, A.}, \bibinfo{author}{Kim, K.}, \bibinfo{author}{Warfield, W.}, \bibinfo{author}{Safronova, M.~S.} \& \bibinfo{author}{Ye, J.}
\newblock \bibinfo{title}{{Clock with 8×10$^{-19}$ Systematic Uncertainty}}.
\newblock \emph{\bibinfo{journal}{Phys. Rev. Lett.}} \textbf{\bibinfo{volume}{133}}, \bibinfo{pages}{023401} (\bibinfo{year}{2024}).

\bibitem{AnnPhys19Thirolf}
\bibinfo{author}{Thirolf, P.~G.}, \bibinfo{author}{Seiferle, B.} \& \bibinfo{author}{von~der Wense, L.}
\newblock \bibinfo{title}{{Improving Our Knowledge on the ${}^{229m}$Th Isomer: Toward a Test Bench for Time Variations of Fundamental Constants}}.
\newblock \emph{\bibinfo{journal}{Ann. Phys.}} \textbf{\bibinfo{volume}{531}}, \bibinfo{pages}{1800381} (\bibinfo{year}{2019}).

\bibitem{peik2021nuclear}
\bibinfo{author}{Peik, E.}, \bibinfo{author}{Schumm, T.}, \bibinfo{author}{Safronova, M.~S.}, \bibinfo{author}{P{\'{a}}lffy, J., A.and~Weitenberg} \& \bibinfo{author}{Thirolf, P.~G.}
\newblock \bibinfo{title}{Nuclear clocks for testing fundamental physics}.
\newblock \emph{\bibinfo{journal}{Quantum Sci. Technol.}} \textbf{\bibinfo{volume}{6}}, \bibinfo{pages}{034002} (\bibinfo{year}{2021}).

\bibitem{Flambaum06}
\bibinfo{author}{Flambaum, V.~V.}
\newblock \bibinfo{title}{{Enhanced Effect of Temporal Variation of the Fine Structure Constant and the Strong Interaction in $^{229}$Th}}.
\newblock \emph{\bibinfo{journal}{Phys. Rev. Lett.}} \textbf{\bibinfo{volume}{97}}, \bibinfo{pages}{092502} (\bibinfo{year}{2006}).

\bibitem{Berengut09}
\bibinfo{author}{Berengut, J.~C.}, \bibinfo{author}{Dzuba, V.~A.}, \bibinfo{author}{Flambaum, V.~V.} \& \bibinfo{author}{Porsev, S.~G.}
\newblock \bibinfo{title}{{Proposed Experimental Method to Determine $\alpha$ Sensitivity of Splitting between Ground and \SI{7.6}{\electronvolt} Isomeric States in ${}^{229}$Th}}.
\newblock \emph{\bibinfo{journal}{Phys. Rev. Lett.}} \textbf{\bibinfo{volume}{102}}, \bibinfo{pages}{210801} (\bibinfo{year}{2009}).

\bibitem{Beeks2024}
\bibinfo{author}{Beeks, K.} \emph{et~al.}
\newblock \bibinfo{title}{{Fine-structure constant sensitivity of the Th-229 nuclear clock transition}}.
\newblock \emph{\bibinfo{journal}{arXiv 2407.17300}}  (\bibinfo{year}{2024}).

\bibitem{ThirolfEPJ24}
\bibinfo{author}{{Thirolf, P. G.}}, \bibinfo{author}{{Kraemer, S.}}, \bibinfo{author}{{Moritz, D.}} \& \bibinfo{author}{{Scharl, K.}}
\newblock \bibinfo{title}{{The thorium isomer ${}^{229m}$Th: review of status and perspectives after more than 50 years of research}}.
\newblock \emph{\bibinfo{journal}{Eur. Phys. J. Spec. Top.}} \bibinfo{pages}{1--19} (\bibinfo{year}{2024}).

\bibitem{hiraki2024}
\bibinfo{author}{Hiraki, T.} \emph{et~al.}
\newblock \bibinfo{title}{{Controlling $^{229}$Th isomeric state population in a VUV transparent crystal}}.
\newblock \emph{\bibinfo{journal}{Nat. Commun.}} \textbf{\bibinfo{volume}{15}}, \bibinfo{pages}{5536} (\bibinfo{year}{2024}).

\bibitem{ThoriumNuclearClock}
\bibinfo{author}{{ThoriumNuclearClock}}.
\newblock \bibinfo{title}{{Thorium nuclear clock website}}.
\newblock \urlprefix\url{https://thoriumclock.eu/}.
\newblock \bibinfo{note}{(accessed on 04.12.2024)}.

\bibitem{schwarz2012cryogenic}
\bibinfo{author}{Schwarz, M.} \emph{et~al.}
\newblock \bibinfo{title}{{Cryogenic linear Paul trap for cold highly charged ion experiments}}.
\newblock \emph{\bibinfo{journal}{Rev. Sci. Instrum.}} \textbf{\bibinfo{volume}{83}}, \bibinfo{pages}{083115} (\bibinfo{year}{2012}).

\bibitem{leopold2019cryogenic}
\bibinfo{author}{Leopold, T.} \emph{et~al.}
\newblock \bibinfo{title}{A cryogenic radio-frequency ion trap for quantum logic spectroscopy of highly charged ions}.
\newblock \emph{\bibinfo{journal}{Rev. Sci. Instrum.}} \textbf{\bibinfo{volume}{90}}, \bibinfo{pages}{073201} (\bibinfo{year}{2019}).

\bibitem{ColdEdgeVibrationInterface}
\bibinfo{author}{{ColdEdge Technologies}}.
\newblock \bibinfo{title}{{Low Vibration Interface -- Owner's Manual}}.
\newblock \bibinfo{howpublished}{Technical Information}.

\bibitem{dubielzig2021ultra}
\bibinfo{author}{Dubielzig, T.} \emph{et~al.}
\newblock \bibinfo{title}{Ultra-low-vibration closed-cycle cryogenic surface-electrode ion trap apparatus}.
\newblock \emph{\bibinfo{journal}{Rev. Sci. Instrum.}} \textbf{\bibinfo{volume}{92}}, \bibinfo{pages}{043201} (\bibinfo{year}{2021}).

\bibitem{Dubielzig_Thesis}
\bibinfo{author}{Dubielzig, T.}
\newblock \emph{\bibinfo{title}{Ultra-low vibration closed-cycle cryogenic surface-electrode ion trap apparatus}}.
\newblock \bibinfo{type}{{PhD thesis}}, \bibinfo{school}{Leibniz Universität Hannover} (\bibinfo{year}{2021}).

\bibitem{ScharlSetup2023}
\bibinfo{author}{Scharl, K.} \emph{et~al.}
\newblock \bibinfo{title}{{Setup for the Ionic Lifetime Measurement of the ${}^{229m}$Th$^{3+}$ Nuclear Clock Isomer}}.
\newblock \emph{\bibinfo{journal}{Atoms}}  (\bibinfo{year}{2023}).

\bibitem{NeumayrThesis}
\bibinfo{author}{Neumayr, J.}
\newblock \emph{\bibinfo{title}{{The buffer-gas cell and the extraction RFQ for SHIPTRAP}}}.
\newblock \bibinfo{type}{{PhD thesis}}, \bibinfo{school}{Ludwig-Maximilians-Universität München} (\bibinfo{year}{2004}).

\bibitem{Neumayr06}
\bibinfo{author}{Neumayr, J.~B.} \emph{et~al.}
\newblock \bibinfo{title}{{Performance of the MLL-IonCatcher}}.
\newblock \emph{\bibinfo{journal}{Rev. Sci. Instrum.}} \textbf{\bibinfo{volume}{77}}, \bibinfo{pages}{065109} (\bibinfo{year}{2006}).

\bibitem{Thielking18}
\bibinfo{author}{Thielking, J.} \emph{et~al.}
\newblock \bibinfo{title}{{Laser spectroscopic characterization of the nuclear-clock isomer ${}^{229m}$Th}}.
\newblock \emph{\bibinfo{journal}{Nature}} \textbf{\bibinfo{volume}{556}}, \bibinfo{pages}{321--325} (\bibinfo{year}{2018}).

\bibitem{Wense18}
\bibinfo{author}{von~der Wense, L.}, \bibinfo{author}{Seiferle, B.} \& \bibinfo{author}{Thirolf, P.~G.}
\newblock \bibinfo{title}{{Towards a ${}^{229}$Th-based nuclear clock}}.
\newblock \emph{\bibinfo{journal}{Meas. Tech.}} \textbf{\bibinfo{volume}{60}}, \bibinfo{pages}{1178--1192} (\bibinfo{year}{2018}).

\bibitem{Eberhardt2028}
\bibinfo{author}{Eberhardt, K.} \emph{et~al.}
\newblock \bibinfo{title}{Actinide targets for fundamental research in nuclear physics}.
\newblock \emph{\bibinfo{journal}{AIP Conf. Proc.}} \textbf{\bibinfo{volume}{1962}}, \bibinfo{pages}{030009} (\bibinfo{year}{2018}).

\bibitem{WakelingThesisWSU}
\bibinfo{author}{Wakeling, M.~A.}
\newblock \bibinfo{title}{{Charge states of ${}^{229m}$Th: Path to finding the half-life}} (\bibinfo{year}{2014}).

\bibitem{Thirolf19}
\bibinfo{author}{Thirolf, P.~G.}, \bibinfo{author}{Seiferle, B.} \& \bibinfo{author}{von~der Wense, L.}
\newblock \bibinfo{title}{The 229-thorium isomer: doorway to the road fromthe atomic clock to the nuclear clock}.
\newblock \emph{\bibinfo{journal}{J. Phys. B}} \textbf{\bibinfo{volume}{52}}, \bibinfo{pages}{203001} (\bibinfo{year}{2019}).

\bibitem{TimoDickel}
\bibinfo{author}{{Dickel, T.}}
\newblock \bibinfo{howpublished}{Justus Liebig Universität Gießen, private communication} (\bibinfo{year}{2019}).

\bibitem{turchette2000heating}
\bibinfo{author}{Turchette, Q.~A.} \emph{et~al.}
\newblock \bibinfo{title}{Heating of trapped ions from the quantum ground state}.
\newblock \emph{\bibinfo{journal}{Phys. Rev. A}} \textbf{\bibinfo{volume}{61}}, \bibinfo{pages}{063418} (\bibinfo{year}{2000}).

\bibitem{deVoe2002experimental}
\bibinfo{author}{DeVoe, R.~G.} \& \bibinfo{author}{Kurtsiefer, C.}
\newblock \bibinfo{title}{{Experimental study of anomalous heating and trap instabilities in a microscopic $^{137}\mathrm{Ba}$ ion trap}}.
\newblock \emph{\bibinfo{journal}{Phys. Rev. A}} \textbf{\bibinfo{volume}{65}}, \bibinfo{pages}{063407} (\bibinfo{year}{2002}).

\bibitem{daniilidis2011fabrication}
\bibinfo{author}{Daniilidis, N.} \emph{et~al.}
\newblock \bibinfo{title}{Fabrication and heating rate study of microscopic surface electrode ion traps}.
\newblock \emph{\bibinfo{journal}{New J. Phys.}} \textbf{\bibinfo{volume}{13}}, \bibinfo{pages}{013032} (\bibinfo{year}{2011}).

\bibitem{haerter2014long}
\bibinfo{author}{H{\"a}rter, A.}, \bibinfo{author}{Kr{\"u}kow, A.}, \bibinfo{author}{Brunner, A.} \& \bibinfo{author}{Hecker~Denschlag, J.}
\newblock \bibinfo{title}{Long-term drifts of stray electric fields in a paul trap}.
\newblock \emph{\bibinfo{journal}{Appl. Phys. B}} \textbf{\bibinfo{volume}{114}}, \bibinfo{pages}{275--281} (\bibinfo{year}{2014}).

\bibitem{Wense15}
\bibinfo{author}{Von~der Wense, L.}, \bibinfo{author}{Seiferle, B.}, \bibinfo{author}{Laatiaoui, M.} \& \bibinfo{author}{Thirolf, P.~G.}
\newblock \bibinfo{title}{{Determination of the extraction efficiency for $^{233}$U source $\alpha$-recoil ions from the MLL buffer-gas stopping cell}}.
\newblock \emph{\bibinfo{journal}{Eur. Phys. J. A}} \textbf{\bibinfo{volume}{51}}, \bibinfo{pages}{29} (\bibinfo{year}{2015}).

\bibitem{HaettnerThesis}
\bibinfo{author}{Haettner, E.}
\newblock \emph{\bibinfo{title}{{A novel radio frequency quadrupole system for SHIPTRAP $\&$ New mass measurements of rp nuclides}}}.
\newblock \bibinfo{type}{{PhD thesis}}, \bibinfo{school}{Justus-Liebig-Universität Gießen} (\bibinfo{year}{2011}).

\bibitem{Haettner2018}
\bibinfo{author}{Haettner, E.} \emph{et~al.}
\newblock \bibinfo{title}{A versatile triple radiofrequency quadrupole system for cooling, mass separation and bunching of exotic nuclei}.
\newblock \emph{\bibinfo{journal}{Nucl. Instrum. Methods Phys. Res. A}} \textbf{\bibinfo{volume}{880}}, \bibinfo{pages}{138--151} (\bibinfo{year}{2018}).

\bibitem{Brubaker1968}
\bibinfo{author}{Brubaker, W.}
\newblock \bibinfo{title}{{An improved quadrupole mass analyser}}.
\newblock \emph{\bibinfo{journal}{Adv. Mass Spectrom.}} \textbf{\bibinfo{volume}{4}}, \bibinfo{pages}{293--299} (\bibinfo{year}{1968}).

\bibitem{Douglas2002}
\bibinfo{author}{Douglas, D.~J.} \& \bibinfo{author}{Konenkov, N.~V.}
\newblock \bibinfo{title}{Influence of the 6th and 10th spatial harmonics on the peak shape of a quadrupole mass filter with round rods}.
\newblock \emph{\bibinfo{journal}{Rapid Commun. Mass Spectrom.}} \textbf{\bibinfo{volume}{16}}, \bibinfo{pages}{1425--1431} (\bibinfo{year}{2002}).

\bibitem{TracoTHV}
\bibinfo{author}{{TRACO Power Group}}.
\newblock \bibinfo{title}{{High Voltage Power Supplies THV Series, Documentation}}.
\newblock \bibinfo{howpublished}{Technical Information}.

\bibitem{NI9264}
\bibinfo{author}{{National Instruments}}.
\newblock \bibinfo{title}{{NI 9264 Datasheet}}.
\newblock \bibinfo{howpublished}{Technical Information}.

\bibitem{Coolinghead}
\bibinfo{author}{Cryogenics Division Precision Equipment Group Sumitomo Heavy~Industries, L.}
\newblock \bibinfo{title}{Shi cryocooler specification -- model: Srp-082b-f70h}.
\newblock \bibinfo{howpublished}{Technical Information} (\bibinfo{year}{2009}).

\bibitem{Wissenberg2025}
\bibinfo{author}{Wissenberg, S.~H.} \emph{et~al.}
\newblock \bibinfo{title}{Noncollinear enhancement resonator with intrinsic pulse synchronization and alignment employing wedge mirrors}.
\newblock \emph{\bibinfo{journal}{Phys. Rev. Res.}} \textbf{\bibinfo{volume}{7}}, \bibinfo{pages}{023071} (\bibinfo{year}{2025}).

\bibitem{PfeifferHiPacePump}
\bibinfo{author}{{Pfeiffer Vacuum GmbH}}.
\newblock \bibinfo{title}{{Specifications -- HiPace 300 H}}.
\newblock \bibinfo{howpublished}{Technical Information}.

\bibitem{Edwardsxds35i}
\bibinfo{author}{{Edwards Japan Limited}}.
\newblock \bibinfo{title}{{Product Data Sheet -- XDS35i Dry Scroll Pumps}}.
\newblock \bibinfo{howpublished}{Technical Information}.

\bibitem{EdwardsiX457}
\bibinfo{author}{{Edwards Japan Limited}}.
\newblock \bibinfo{title}{{Instruction Manual -- STP Series Turbomolecular Pumps -- STP-iX457/iXU457 Series}}.
\newblock \bibinfo{howpublished}{Technical Information}.

\bibitem{EdwardsSTP603}
\bibinfo{author}{{Edwards Japan Limited}}.
\newblock \bibinfo{title}{{Turbo Molecular Pump STP-603/1003 series Specification}}.
\newblock \bibinfo{howpublished}{Technical Information}.

\bibitem{AgilentPumpe}
\bibinfo{author}{{Agilent Technologies}}.
\newblock \bibinfo{title}{{Specifications -- Agilent TwisTorr 84 FS}}.
\newblock \bibinfo{howpublished}{Technical Information}.

\bibitem{PfeifferPBR260}
\bibinfo{author}{{Pfeiffer Vacuum GmbH}}.
\newblock \bibinfo{title}{{Operating Instructions PBR 260 Compact FullRange BA Gauge}}.
\newblock \bibinfo{howpublished}{Technical Information}.

\bibitem{PfeifferHiPace80neo}
\bibinfo{author}{{Pfeiffer Vacuum GmbH}}.
\newblock \bibinfo{title}{{Operating Instructions -- HiPace 80 Neo Turbopump}}.
\newblock \bibinfo{howpublished}{Technical Information}.

\bibitem{dahl2000simion}
\bibinfo{author}{Dahl, D.~A.}
\newblock \bibinfo{title}{{SIMION for the personal computer in reflection}}.
\newblock \emph{\bibinfo{journal}{Int. J. Mass Spectrom.}} \textbf{\bibinfo{volume}{200}}, \bibinfo{pages}{3--25} (\bibinfo{year}{2000}).

\bibitem{WenseThesis}
\bibinfo{author}{von~der Wense, L.}
\newblock \emph{\bibinfo{title}{{On the direct detection of ${}^{229m}$Th}}}.
\newblock \bibinfo{type}{{PhD thesis}}, \bibinfo{school}{Ludwig-Maximilians-Universität München} (\bibinfo{year}{2016}).

\bibitem{removille2009trapping}
\bibinfo{author}{Removille, S.} \emph{et~al.}
\newblock \bibinfo{title}{{Trapping and cooling of Sr$^{+}$ions: strings and large clouds}}.
\newblock \emph{\bibinfo{journal}{J. Phys. B}} \textbf{\bibinfo{volume}{42}}, \bibinfo{pages}{154014} (\bibinfo{year}{2009}).

\bibitem{jung2017all}
\bibinfo{author}{Jung, K.} \emph{et~al.}
\newblock \bibinfo{title}{{All-diode-laser cooling of Sr$^{+}$ isotope ions for analytical applications}}.
\newblock \emph{\bibinfo{journal}{Jpn. J. Appl. Phys.}} \textbf{\bibinfo{volume}{56}}, \bibinfo{pages}{062401} (\bibinfo{year}{2017}).

\bibitem{gallagher1967oscillator}
\bibinfo{author}{Gallagher, A.}
\newblock \bibinfo{title}{{Oscillator Strengths of Ca II, Sr II, and Ba II}}.
\newblock \emph{\bibinfo{journal}{Phys. Rev.}} \textbf{\bibinfo{volume}{157}}, \bibinfo{pages}{24--30} (\bibinfo{year}{1967}).

\bibitem{ThoriumnuclearClockPI}
\bibinfo{author}{Peik, E.} \emph{et~al.}
\newblock \bibinfo{title}{Nuclear clocks for testing fundamental physics}.
\newblock \emph{\bibinfo{journal}{Quantum Sci. Technol.}} \textbf{\bibinfo{volume}{6}}, \bibinfo{pages}{034002} (\bibinfo{year}{2021}).

\bibitem{zitzer2024sympathetic}
\bibinfo{author}{Zitzer, G.} \emph{et~al.}
\newblock \bibinfo{title}{{Sympathetic cooling of trapped ${\mathrm{Th}}^{3+}$ alpha-recoil ions for laser spectroscopy}}.
\newblock \emph{\bibinfo{journal}{Phys. Rev. A}} \textbf{\bibinfo{volume}{109}}, \bibinfo{pages}{033116} (\bibinfo{year}{2024}).

\bibitem{schmoeger2015coulomb}
\bibinfo{author}{Schmöger, L.} \emph{et~al.}
\newblock \bibinfo{title}{Coulomb crystallization of highly charged ions}.
\newblock \emph{\bibinfo{journal}{Science}} \textbf{\bibinfo{volume}{347}}, \bibinfo{pages}{1233--1236} (\bibinfo{year}{2015}).

\bibitem{schmoeger2017kalte}
\bibinfo{author}{Schmöger, L.}
\newblock \emph{\bibinfo{title}{Kalte hochgeladene Ionen für Frequenzmetrologie}}.
\newblock \bibinfo{type}{{PhD thesis}}, \bibinfo{school}{Ruprech-Karls-Universität Heidelberg}, \bibinfo{address}{Heidelberg} (\bibinfo{year}{2017}).

\bibitem{pagano2019cryogenic}
\bibinfo{author}{Pagano, G.} \emph{et~al.}
\newblock \bibinfo{title}{Cryogenic trapped-ion system for large scale quantum simulation}.
\newblock \emph{\bibinfo{journal}{Quantum Sci. Technol.}} \textbf{\bibinfo{volume}{4}}, \bibinfo{pages}{014004} (\bibinfo{year}{2018}).

\bibitem{repp2012pentrap}
\bibinfo{author}{Repp, J.} \emph{et~al.}
\newblock \bibinfo{title}{{PENTATRAP: a novel cryogenic multi-Penning-trap experiment for high-precision mass measurements on highly charged ions}}.
\newblock \emph{\bibinfo{journal}{Appl. Phys. B}} \textbf{\bibinfo{volume}{107}}, \bibinfo{pages}{983--996} (\bibinfo{year}{2012}).

\bibitem{sellner2017improved}
\bibinfo{author}{Sellner, S.} \emph{et~al.}
\newblock \bibinfo{title}{{Improved limit on the directly measured antiproton lifetime}}.
\newblock \emph{\bibinfo{journal}{New J. Phys.}} \textbf{\bibinfo{volume}{19}}, \bibinfo{pages}{083023} (\bibinfo{year}{2017}).

\bibitem{micke2019closed}
\bibinfo{author}{Micke, P.} \emph{et~al.}
\newblock \bibinfo{title}{{Closed-cycle, low-vibration 4 K cryostat for ion traps and other applications}}.
\newblock \emph{\bibinfo{journal}{Rev. Sci. Instrum.}} \textbf{\bibinfo{volume}{90}}, \bibinfo{pages}{065104} (\bibinfo{year}{2019}).

\bibitem{Godun2014}
\bibinfo{author}{Godun, R.~M.} \emph{et~al.}
\newblock \bibinfo{title}{{Frequency Ratio of Two Optical Clock Transitions in $^{171}{\mathrm{Yb}}^{+}$ and Constraints on the Time Variation of Fundamental Constants}}.
\newblock \emph{\bibinfo{journal}{Phys. Rev. Lett.}} \textbf{\bibinfo{volume}{113}}, \bibinfo{pages}{210801} (\bibinfo{year}{2014}).

\end{thebibliography}

\end{document}